%% file: fuzzy3d.tex
\documentclass[12pt]{article}

\usepackage{epsfig,multicol,multirow}
\usepackage{amsmath,subfigure,latexsym,amssymb}

\def\lsi{\raise0.3ex\hbox{$<$\kern-0.75em\raise-1.1ex\hbox{$\sim$}}}
\def\gsi{\raise0.3ex\hbox{$>$\kern-0.75em\raise-1.1ex\hbox{$\sim$}}}
\def\backder{\raise1.4ex\hbox{$\leftarrow$\kern-0.75em\raise-1.4ex\hbox{$\partial$}}}
\def\forder{\raise1.4ex\hbox{$\rightarrow$\kern-0.75em\raise-1.4ex\hbox{$\partial$}}}

\newcommand{\lsim}{\mathop{\lsi}}
\newcommand{\gsim}{\mathop{\gsi}}

\newcommand{\be}{\begin{equation}}
\newcommand{\ee}{\end{equation}}
\newcommand{\nn}{\nonumber}
\newcommand{\bea}{\begin{eqnarray}}
\newcommand{\eea}{\end{eqnarray}} 

\newcommand{\la}{\langle}
\newcommand{\ra}{\rangle}
\newcommand{\uno}{1 \!\! 1}

\newcommand{\R}{{\kern+.25em\sf{R}\kern-.78em\sf{I} \kern+.78em\kern-.25em}}
\newcommand{\RR}{{\kern+.25em\sf{R}\kern-.6em\sf{I} \kern+.6em\kern-.25em}}
\newcommand{\N}{{\kern+.25em\sf{N}\kern-.78em\sf{I} \kern+.78em\kern-.25em}}
\newcommand{\C}{{\kern+.25em\sf{C}\kern-.50em\sf{I} \kern+.50em\kern-.25em}}

\makeatletter
\@addtoreset{equation}{section}
\makeatother

\newcommand{\ijtp}{\emph{Int.\ J.\ Mod.\ Phys.}}
\newcommand{\cqg}{\emph{Class.\ and Quant.\ Grav.}}
\newcommand{\cmp}{\emph{Commun.\ Math.\ Phys.}}

\newcommand{\jhep}{\emph{JHEP}}

\begin{document}

\input{title2}

\newpage

\tableofcontents


\section{Introduction}

\input{intro2}

\section{The fuzzy sphere formulation of the 3d $\lambda \phi^{4}$ model}

\input{formulation2}

\section{Determination of the phase diagram}

\input{phases2}

\section{The scaling of the phase transitions}

\input{scaling}

\section{Confrontation with related models}

\input{redu2}

\section{A conjecture about the large $N$ limit}

\input{conj}

\section{Conclusions}

\input{conclu2}

\appendix

\section{Technical aspects of the simulation}

\input{simu2}

\input{acknow2}

\input{refs2}
\end{document}

%% file: title2.tex
\begin{flushright}
DESY-07-206 
\end{flushright}

\vspace*{6mm}

\begin{center}

{\Large\bf Probing the fuzzy sphere regularisation} \\
\vspace*{5mm}
{\Large\bf in simulations of the 3d $\lambda \phi^{4}$ model}

\vspace*{15mm}

Julieta Medina$^{\rm \, a}$, Wolfgang Bietenholz$^{\rm \, b}$ and 
Denjoe O'Connor$^{\rm \, c}$ \\

\vspace*{5mm}

$^{\rm a}$ 
Ciencias B\'{a}sicas, UPIITA \\
Instituto Polit\'{e}cnico Nacional (IPN) \\
Av.\ Inst.\ Polit\'{e}cnico 2508 \\
C.P.\ 07340 M\'{e}xico D.F., M\'{e}xico \\

\vspace*{3mm}

$^{\rm b}$ John von Neumann Institut (NIC) \\
DESY Zeuthen, Platanenallee 6 \\
D-15738 Zeuthen, Germany \\

\vspace*{3mm}

$^{\rm c}$ Dublin Institute for Advanced Studies (DIAS) \\
10, Burlington Road, Dublin 4, Ireland

\end{center}

\vspace*{1cm}

We regularise the 3d $\lambda \phi^{4}$ model by discretising
the Euclidean time and representing the spatial part on a fuzzy
sphere. The latter involves a truncated expansion of the
field in spherical harmonics. This yields a numerically
tractable formulation, which constitutes an unconventional alternative
to the lattice. In contrast to the 2d version, the radius $R$
plays an independent r\^{o}le. We explore the phase diagram in terms 
of $R$ and the cutoff, as well as
the parameters $m^{2}$ and $\lambda$. Thus we 
identify the phases of disorder, uniform order and non-uniform order.
We compare the result to the phase diagrams of the 3d model
on a non-commutative torus, and of the 2d model on a fuzzy sphere.
Our data at strong coupling reproduce accurately the
behaviour of a matrix chain, which corresponds to the $c=1$--model
in string theory. This observation enables a conjecture about
the thermodynamic limit.

%% file: intro2.tex
A variety of approaches to the regularisation of quantum field theory exist.
Dimensional regularisation \cite{dimreg}
is most popular in the framework of perturbation theory.
In order to overcome the limitations of a perturbative
expansion, however, a regularisation should restrict the
formulation to a finite set of degrees of freedom. If the
Euclidean action is real and bounded from below, a model can then be 
treated numerically as  a statistical system. This method provides
in many cases the only access to observables beyond perturbation
theory or semi-classical approximations.

From a general perspective, field theoretic models start
from some algebra ${\cal A}$ for functions on a manifold ${\cal M}$,
and a differential operator ${\cal D}$ with its Hilbert space ${\cal H}$.
The standard approach for non-perturbative studies discretises the 
manifold to a lattice. Then the degrees of freedom to work with
are usually the field variables on the lattice sites or links 
(see e.g.\ Ref.\ \cite{MM}). As an alternative, also Monte Carlo 
simulations employing 
the fields at discrete momenta have been suggested 
\cite{psimu}, though much less 
explored.\footnote{Of course, the standard Hybrid Monte Carlo algorithm
for dynamical fermions includes a Langevin ingredient in momentum space,
but in that case the basic regularisation is nevertheless a space-time 
lattice.} 
In both cases ${\cal M}$ is reduced to a finite lattice.

Generally the goal is to approximate
a triple $({\cal A}, {\cal H}, {\cal D})$ \cite{Connes}.
This might be achieved in quite abstract ways, but
in practice a physical picture for the regularised system 
is a useful guide-line. In the lattice formulation one
approximates the entire triple. The algebra is approximated
by a commutative algebra  on a lattice of points which
discretise ${\cal M}$, while the differential operator is
obtained by a finite difference approximation to ${\cal D}$,
and the Hilbert space is adapted to this operator.

Here we are concerned with an alternative scheme, which is
endowed with a physical picture on the regularised level as well.
Instead of the discrete eigenvalues
of the space-time or momentum coordinates, we now deal with
{\em angular momentum coordinates.} 
To this end the fields are wrapped on a sphere and expanded 
in spherical harmonics.
A related idea 
occurred already in an early construction of
a non-commutative space, which added an extra dimension
and preserved 5d Lorentz symmetry \cite{Sny}. This method benefits from
the natural discretisation of angular momentum space in quantum
physics, but a 
cutoff still has to be imposed. In the interpretation of the angular momenta
as spherical coordinates, the cutoff renders the sphere fuzzy.
These coordinates are embedded in matrices, which 
are Hermitian in the case of the neutral scalar field 
to be considered here.

The concept of a fuzzy sphere regularisation has been established
in Refs.\ \cite{fuzzy}. The Laplace-Beltrami operator ${\cal D}^{2}$
is given in terms of angular momentum operators, which are expressed 
by $N$-dimensional irreducible representations of $SU(2)$.
The algebra ${\cal A}$ is expanded in
the polarisation tensors, which are matrix analogues of the
spherical harmonics \cite{polten}.
In contrast to the lattice, this regularisation does not
explicitly break the space symmetries.
Its analytic properties have been studied
extensively in recent years, but the applicability
in numerical simulations is less explored. The questions are if such
simulations are feasible and to what kind of limits the
measured observables can be extrapolated. So far the 2d 
$\lambda \phi^{4}$ model has been investigated in this respect
\cite{2dfuzzy}.\footnote{Further numerical studies on the
fuzzy sphere address $U(1)$ gauge theory \cite{BadDen}.}
Here we extend this study to three dimensions,
where the spatial plane is mapped onto a fuzzy sphere, and the
Euclidean time is lattice discretised. This extension entails
qualitative differences, which are essential
in view of the prospects of proceeding to four dimensions. 
In particular the radius $R$ of the sphere plays an independent 
r\^{o}le (it cannot be absorbed by simple rescaling).
The recovery of a flat space without truncation requires the 
limits $N,\, R \to \infty$.

Section 2 presents the fuzzy sphere formulation of the
$\lambda \phi^{4}$ model, along with suitable
order parameters. The identification of the phase transitions
in the $(\lambda , m^{2})$-plane is described in Section 3.
Section 4 discusses the scaling of the phase transition
lines in terms of $N$ and $R$.
Section 5 compares our results to
the phase diagram of the corresponding 2d model on a fuzzy sphere,
and to the model on a 3d non-commutative torus.
We also demonstrate that our data at strong coupling 
agree with the behaviour of
a matrix chain model, which attracted interest in string theory.
This observation allows for a conjecture about the large $N$
extrapolation, as we point out in Section 6.
Our results are summarised in Section 7, and
technicalities of the simulation are added in an Appendix.

A synopsis of this work has been anticipated in a proceeding 
contribution \cite{Lat05}, and details are presented 
in a Ph.D.\ thesis \cite{Julieta}.

%% file: formulation2.tex
\subsection{Regularisation}

In this subsection we specify the regularisation that we used
in our simulations. The theory to
be regularised is the $\lambda \phi^{4}$ model in 3 dimensions,
where we assume periodic boundary conditions in the Euclidean time
$t\,$, and 
the space is taken as a sphere in $\R^{3}$.
Thus the action reads
\be
S [ \phi ] = \int_{0}^{T} dt \int d^{3}x \, \delta (\vec x^{\, 2} - R^{2})
\ \Big[ \frac{1}{2} \phi \Big( - \partial_{t}^{2} + 
\frac{ {\cal L}^{2}}{R^{2}}\Big)
\phi + \frac{m^{2}}{2} \phi^{2} + \frac{\lambda}{4} \phi^{4} \Big] \ .
\ee
$T$ is the temporal periodicity and $R$ is the radius of the
sphere, which we are going to denote by $S_{R}^{2}\,$.
$\phi (t, \vec x) \in \R$ is a scalar field and ${\cal L}^{2} =
\sum_{i=1}^{3} {\cal L}_{i}^{2} \, $, ${\cal L}_{i}$ being the 
angular momentum components. \\

For the regularisation in time we introduce $N_{t}$ equidistant
sites and replace $\partial_{t}$ by the standard lattice
derivative.  We are going to use lattice units, 
i.e.\ we set $N_{t} = T$.
Thus a configuration is given by a set
$\phi_{t}(\vec x)\,$, $t =1, \dots , N_{t}\,$.\\

Our regularisation of the sphere $S_{R}^{2}$ is less standard, but
it also relies on a concept established in the literature \cite{fuzzy}.
The coordinates $x_{i}$ are replaced by operators
$X_{i}$, which still obey the constraint
\be
\sum_{i=1}^{3} X_{i}^{2} = R^{2} \cdot \uno \ .
\ee
A truncation to a maximal angular momentum $\ell_{\rm max}$ means that
the operators $X_{i}$ take the form of $N \times N$ matrices with
$N = \ell_{\max}+1$,
\be  \label{fuzzycor}
X_{i} = \frac{2R}{\sqrt{N^{2} -1}} \, L_{i} \ , \qquad
( X_{i} \in {\rm Mat}_{N}) \ .
\ee
The $L_{i}$ are generators in an $N$-dimensional
irreducible representation of $SU(2)$. These
coordinate operators do not commute,
\be
[ X_{i}, X_{j}] = {\rm i} \, \frac{2R}{\sqrt{N^{2} -1}} \,
\epsilon_{ijk} X_{k} \ .
\ee
Thus they cannot describe sharp points; the sphere becomes
{\em fuzzy}.

A scalar field, which can be expressed as a power series in the
coordinates, now turns into an expansion in the operators
$X_{i}$ (at some fixed time site $t$). 
Thus this formulation represents the field by
$N \times N$ matrices $\Phi_{t}$. 
In particular for the neutral scalar field
these matrices are Hermitian.
Its spatial derivatives are given as commutators,
$\partial_{i} \phi_{t}(\vec x ) \to {\rm i} \, 
[ L_{i}, \Phi_{t}]\,$.

In summary, the recipe for the regularisation from a sharp to
a fuzzy sphere involves the replacements
\bea
x_{i} \in S_{R}^{2} & \rightarrow & X_{i} \in {\rm Mat}_{N} \nn \\
\phi_{t}(\vec x ) \in C^{\infty}(S_{R}^{2}) & \rightarrow &
\Phi_{t} \in {\rm Mat}_{N} \qquad {\rm (Hermitian)} \nn \\
{\cal L}^{2} \phi_{t}(\vec x ) & \rightarrow & \hat {\cal L}^{2} \Phi_{t} 
:= \sum_{i=1}^{3} \, [ L_{i} , [ L_{i}, \Phi_{t} ]] \nn \\
\frac{1}{4 \pi R^{2}} \int_{S_{R}^{2}} d \Omega \,
\phi_{t}(\varphi , \vartheta ) 
& \rightarrow & \frac{1}{N} {\rm Tr} (\Phi_{t} ) \ ,
\eea
where the last relation preserves the normalisation
($d \Omega = R^{2} \sin \vartheta \, d \vartheta \, d \varphi$). \\

We implement these transitions at each discrete time
site $t$. This leads to field configurations given by
$\Phi = \{ \Phi_{1} , \Phi_{2} ,
\dots , \Phi_{N_{t}} \}$, and to the action
\bea
S [ \Phi ] &=& \frac{4 \pi R^{2}}{N} {\rm Tr} \, \Big[  
\sum_{t=1}^{N_{t}} \Big\{ \frac{1}{2} (\Phi_{t+1} - \Phi_{t})^{2}
+ \frac{1}{2R^{2}} \Phi_{t} \, \hat {\cal L}^{2} \, \Phi_{t} \nn \\
&& \hspace*{2.8cm} 
+ \frac{m^{2}}{2} \Phi_{t}^{2} + \frac{\lambda}{4} \Phi_{t}^{4}
\Big\} \Big] \ .  \label{action}
\eea
In this regularised form the functional integral reduces to an 
integration over the independent elements of the Hermitian matrices.
This is in fact tractable in Monte Carlo simulations;
note also that the action (\ref{action}) is real.

A qualitative difference from the 2d model on a fuzzy sphere
(which corresponds to our model on a single time site)
is that the radius $R$ plays an independent r\^{o}le;
it cannot be absorbed in the coupling constants.

A virtue of this approach --- compared to the usual space discretisation 
--- is that continuous spatial rotational symmetry persists on the
regularised level. The fuzzy sphere is rotated by the
adjoint action of an element $U \in SU(2)$ in the $N$-dimensional 
irreducible representation,
\be
\vec X \ \to U^{\dagger} \vec X U = {\cal R} \vec X \ , \qquad
\Phi_{t} \ \to \  U^{\dagger} \Phi_{t} U \ ,
\ee
where $U$ can be written in the form $U = \exp ({\rm i} 
\vec \omega \vec L \, )\,$, and ${\cal R} \in SO(3)$.
A global rotation in all time sites leaves the action 
(\ref{action}) invariant.

This virtue may prove particularly powerful in cases where continuous 
rotational and translational symmetry (which we obtain in the large
$R$ limit) play a central r\^{o}le, such as supersymmetric 
models.\footnote{Literature on supersymmetric systems on a
fuzzy sphere exists regarding the theoretical basis \cite{SUSY}
and first simulations \cite{SUSYsim}, though there are many outstanding
issues in that field. Another important point in this context is that
--- in addition to the space symmetries --- also chiral
symmetry is intact on the fuzzy sphere, without a fermion
doubling problem \cite{BalIm}.}

The question how profitable that symmetry ultimately is has to
be investigated based on non-perturbative results. A prerequisite is
a controlled large $N$ limit, and testing this property is a goal
of the current work. It should also illuminate the 
status of possible pitfalls. In particular, the non-commutativity
of the operators $X_{i}$ implies a non-locality of the interaction
in the regularised model. We are going to see that this property can
indeed affect the thermodynamic limit.\footnote{The impact of a 
non-local regularisation on the continuum limit
is intensively discussed in the lattice community
(see e.g.\ Refs.\ \cite{Creutz}) 
in particular in the light of recent
large-scale QCD simulations with ``rooted staggered fermions''. 
\label{stagg}}
A further goal is to elaborate links of the observed universality 
class to other models of interest.

\subsection{Observables}

Now we introduce the observables to be measured numerically.
For this purpose we first perform a field decomposition, which is
compatible with the rotational symmetry.

The original field $\phi$ can be decomposed in the basis of spherical
harmonics $Y_{\ell m}$ on $S_{R}^{2} \,$,
\be
\phi (t, \varphi, \vartheta ) = \sum_{\ell =0}^{\infty} 
\sum_{m = -\ell}^{\ell} c_{\ell m}(t) Y_{\ell m} (\varphi, \vartheta ) \ .
\ee
In full analogy, the regularised space ${\rm Mat}_{N}$ has a basis
consisting of the {\em polarisation tensors} $\hat Y_{\ell m}$,
see e.g.\ Ref.\ \cite{polten}. For $\ell = 0, \dots , N-1$, 
$m = -\ell , \dots \ell$ these are $N^{2}$ matrices with the
characteristic properties
\bea
\frac{4\pi}{N} {\rm Tr} ( \hat Y^{\dagger}_{\ell ' m'}
\hat Y_{\ell m}) &=& \delta_{\ell ' \ell} \, \delta_{m' m} \ , 
\nn \\
\hat Y^{\dagger}_{\ell (-m)} &=& (-1)^{m} \hat Y_{\ell m} \ , \nn \\
\hat {\cal L}^{2} \hat Y_{\ell m} &=& \ell (\ell +1) 
\hat Y_{\ell m} \ .
\eea
Their construction is reviewed in Ref.\ \cite{Julieta}.
The leading examples are
\bea
\hat Y_{00} &=& \frac{1}{\sqrt{4 \pi}} \, \uno_{N} \nn \\
\hat Y_{10} &=& \sqrt{\frac{3}{\pi (N^{2}-1)}} \, L_{3} \ , 
\quad \hat Y_{1\pm 1} = {\rm i} \, \sqrt{\frac{3}{2 \pi (N^{2}-1)}} \,
L_{\pm} \nn \\
{\rm with} && ( L_{3})_{ij} = \frac{1}{2} (N + 1 -2 i) \,
\delta_{ij} \nn \\
&& ( L_{\pm})_{ij} = ( L_{1} \pm {\rm i} L_{2})_{ij}
= \left\{ \begin{array}{c} 
\sqrt{i (N-i)} \, \delta_{i+1,j} \\
\sqrt{j (N-j)} \, \delta_{i-1,j} \end{array} \right. .  \nn
\eea
We perform this decomposition in each time site,
\be
\Phi_{t} = \sum_{\ell =0}^{N-1} \sum_{m = -\ell}^{\ell}
c_{\ell m} (t) \hat Y_{\ell m} \ ,
\ee
so that $\Phi_{t}$ is fixed by the $N^{2}$ coefficients
\be  \label{clm}
c_{\ell m} (t) = \frac{4 \pi}{N} {\rm Tr} \Big( \hat Y_{\ell m}^{\dagger}
\Phi_{t} \Big) \ .
\ee
In this work we consider the time averaged terms
\be  \label{timeav}
\bar \Phi := \frac{1}{N_{t}} \sum_{t} \Phi_{t} \ , \quad
\bar c_{\ell m} := \frac{1}{N_{t}} \sum_{t} c_{\ell m} (t) \ ,
\ee
which are related as
\be  \label{ctimeav}
\bar c_{00} = \frac{\sqrt{4 \pi}}{N} {\rm Tr} ( \bar \Phi ) \ , \quad 
\bar c_{1m} = \frac{4 \pi}{N} {\rm Tr} 
\Big( \hat Y_{1 m}^{\dagger} \bar \Phi \Big) \ , \quad {\rm etc.}
\ee
We further introduce the quantities
\bea  \label{phil}
\varphi_{\ell}^{2} &:=& \sum_{m=- \ell}^{\ell}
| \bar c_{\ell m} |^{2} \ , 
\quad  \varphi_{\ell} := \sqrt{\varphi_{\ell}^{2}} \ , 
\nn \\
\Vert \bar \Phi \Vert ^{2} &:=& \sum_{\ell =0}^{\infty} \varphi_{\ell}^{2}
= \frac{4 \pi}{N} {\rm Tr} ( \bar \Phi ^{2}) \ .
\eea
We are going to explore the phase diagram by measuring in particular
the order parameters
\be  \label{ordpar}
\langle \varphi_{0} \rangle \quad {\rm and} \quad 
\langle \varphi_{1} \rangle \ . 
\ee
Based on the magnitudes of these expectation values we distinguish
three phases, as we specify in Table \ref{phasetab}. 
Equivalent tools were applied before in
investigations of the 2d model on a fuzzy sphere \cite{2dfuzzy}.
\begin{table}
\begin{center}
\begin{tabular}{|c|c|}
\hline
phase &   \\
\hline
\hline
disordered         & 
$\la \varphi_{0} \ra \approx \la \varphi_{1} \ra \approx 0$ \\
\hline
uniform ordered    &
$\la \varphi_{0} \ra \gg \la \varphi_{1} \ra \approx 0$ \\
\hline
non-uniform ordered & 
$\la \varphi_{1} \ra \gg \la \varphi_{0} \ra \approx 0$ \\
\hline
\end{tabular}
\caption{\emph{The respective magnitudes of the order parameters
$\la \varphi_{0} \ra$ and $ \la \varphi_{1} \ra$
for the three phases that we observed in our results of
Section 3 and 4. Here
the terms in definition (\ref{phil}) are approximately related
as $\langle \Vert \bar \Phi \Vert ^{2} \rangle \approx 
\langle \varphi_{0}^{2} \rangle + \langle \varphi_{1}^{2} \rangle
$, so that the higher contributions
$\langle \varphi_{\ell > 1}^{2} \rangle$ are small.
In general one could distinguish more complicated
ordering structures too. They occur at the strong
couplings, to be addressed in Section 5.}}
\label{phasetab}
\end{center}
\end{table}

\begin{itemize}

\item In the {\em disordered phase} $\la \varphi_{\ell} \ra \approx 0$
holds for all $\ell$. The angular mode decomposition does not detect
any contribution that could indicate a spontaneous breaking of the
rotational symmetry on the sphere.

\item The {\em uniform ordered phase} is characterised by
$ \la \Vert \bar \Phi \Vert ^{2} \ra \approx 
\la \varphi_{0}^{2} \ra \gg 0\,$,
i.e.\  the zero mode contributes significantly, whereas
higher modes are suppressed. This phase corresponds
to the spontaneous magnetisation in a ferromagnet.

\item In the {\em non-uniform ordered phase} a non-zero mode
condenses, which leads to the relation
$$ \la \Vert \bar \Phi \Vert^{2} \ra 
\gg \la \varphi_{0}^{2} \ra \approx 0 \ .
$$ 
In this case the rotational symmetry of the sphere
is spontaneously broken.
For the settings to be explored in Sections 3 and 4, this is manifest 
by a dominant contribution for $\ell =1$ :
$\la \Vert \bar \Phi \Vert ^{2} \ra \approx \la \varphi_{1}^{2} \ra \gg 0$,
$\la \varphi_{\ell \neq 1}^{2} \ra \approx 0 \, $.\\
In the case of strong coupling the non-uniform ordered phase
is dominated by the condensation of higher modes, 
$\la \varphi_{\ell} \ra  \gg
\la \varphi_{1} \ra \approx \la \varphi_{0} \ra \approx 0$
for some $\ell >1$.\\
The general order parameter for this phase reads
$\la \Vert \bar \Phi \Vert ^{2} - \varphi_{0}^{2} \ra $.

\end{itemize}

For a precise identification of the phase transition lines,
we also consider the susceptibility-type observables\footnote{In the
following we will refer to them simply as ``susceptibilities''.}
\be  \label{sus}
\chi_{\ell} := \la \varphi_{\ell}^{2} \ra - 
\la \varphi_{\ell} \ra ^{2} \ ,
\ee
which display peaks at the corresponding phase transitions.

To further substantiate the measurement of the phase diagram
we take thermodynamic quantities into consideration as well,
in particular the internal energy $U$ and the specific heat $C$,
\be \label{sheat}
U = \la S \ra \ , \quad C = \la S^{2} \ra - \la S \ra^{2} \ .
\ee
A peak in $C\,$, and in one of the susceptibilities $\chi_{\ell}\,$,
indicates a (regularised) second order phase transition.

Note that the non-uniform ordered phase is specific to the 
{\em fuzzy} sphere; it does not occur in a regularisation
on a sharp sphere, or
--- generally speaking --- on commutative spaces.
In the flat non-commutative space, such a phase was
predicted in Ref.\ \cite{GuSo} for the $\lambda \phi^{4}$
model in $3$ and $4$ dimensions, as a consequence of the
notorious mixing of ultraviolet and infrared singularities
(UV/IR mixing). For $d=4$ arguments involving the renormalisation 
group \cite{CW} and an effective action \cite{CZ} were added. 
In $d=3$ this behaviour could in fact be demonstrated 
numerically \cite{BHN} by means of lattice simulation results, 
which were extrapolated to a simultaneous UV and IR limit while
keeping the non-commutativity constant (``double scaling
limit''). We will discuss the relation between that result
and the system studied here in Subsection 5.2.

These three phases, including the phase
of non-uniform order, were also observed numerically on the
fuzzy sphere without time direction \cite{2dfuzzy},
in agreement with theoretical considerations 
\cite{fuzzyphi4theo1,fuzzyphi4theo2}.
Similarly this exotic phase was found in lattice studies
of the 2d non-commutative plane \cite{aristocats,BHN}.
However, in that case a double scaling limit has not been
worked out so far, hence the existence of this phase
in the continuous plane is an open question.\footnote{Due to the 
non-locality it is not ruled out by the Mermin-Wagner Theorem.
Still Ref.\ \cite{GuSo} does not expect this phase (for a charged
scalar field) in $d=2$, based on an extension of this Theorem 
to a related effective action with an unusual kinetic term.
For a neutral scalar field Ref.\ \cite{CZ2} arrives at the 
opposite conclusion.}

Here we reconsider the 3d model. However, the spatial part is not 
accommodated on a non-commutative plane but on a fuzzy sphere, 
as we pointed out before.
In addition our main interest refers to the extrapolation
to a commutative limit --- in contrast to double scaling limit
addressed in Ref.\ \cite{BHN} --- in view of the possibility
of using the fuzzy sphere as a regularisation scheme for ordinary
(i.e.\ commutative) field theory.

We performed all our simulations at
\be  \label{convent}
N_{t}=N \ ,
\ee
so that the system has the same number of degrees of freedom
in the temporal and in the spatial directions.

%% file: phases2.tex
To explore the phase diagram we fixed some value of
$\lambda$ and varied $m^{2}$ searching for a phase transition.
Decreasing $m^{2}$ is analogous to lowering the
temperature in statistical mechanics. In all settings we could
identify a critical value $m_{c}^{2} < 0$. For $m^{2} > m_{c}^{2}$
we are in the {\em disordered phase} (corresponding to high 
temperature), whereas $m^{2} < m_{c}^{2}$ gives rise to the dominance of
some ordering, and therefore  spontaneous symmetry breaking.
For small values of $\lambda$ this order is {\em uniform}
(like the spontaneous magnetisation of a ferromagnet), but for larger 
$\lambda$ it becomes {\em non-uniform} (some kind of staggered order), 
cf.\ Section 2.

\begin{figure}[h!]
\vspace*{-3mm}
\begin{center}
\includegraphics[angle=270,width=.5\linewidth]{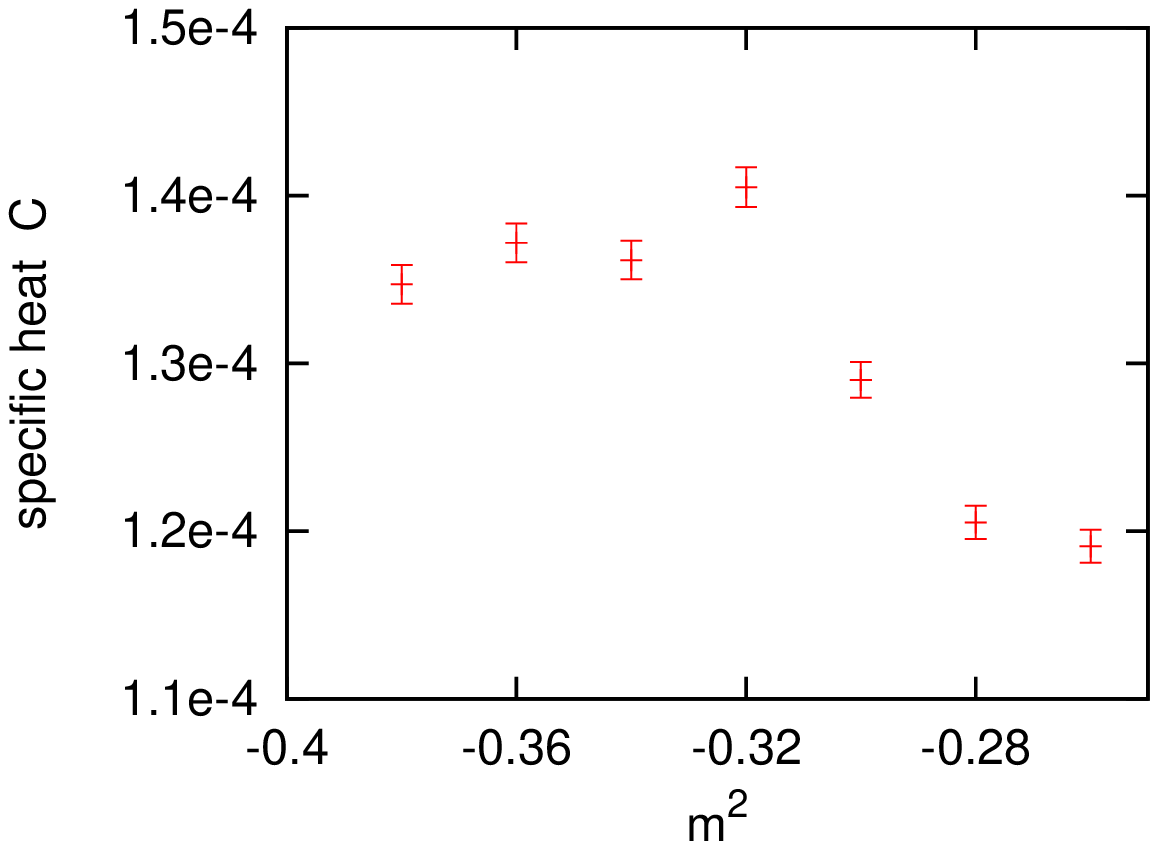}
\hspace*{-3mm}
\includegraphics[angle=270,width=.5\linewidth]{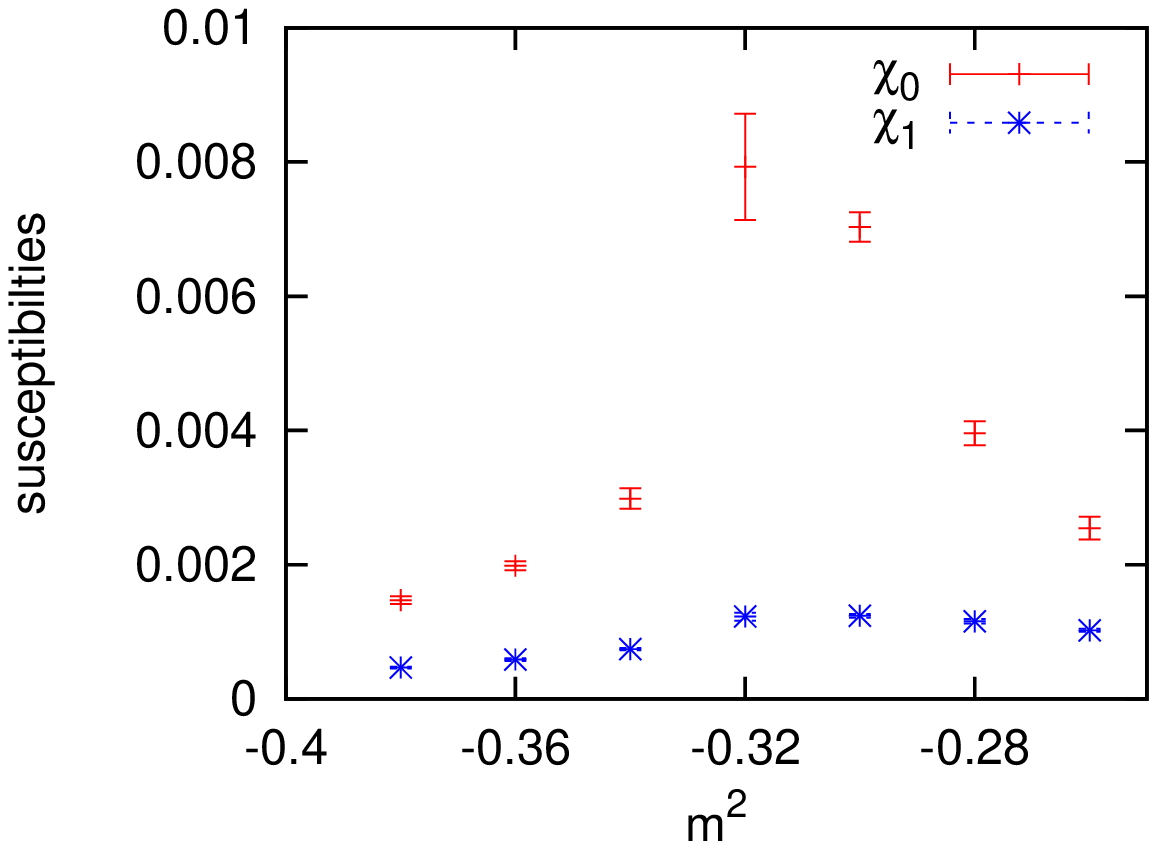}
\vspace*{-5mm} \\
\end{center}
\caption{{\it The specific heat $C$ (on the left) and the susceptibilities
$\chi_{0}$ and $\chi_{1}$ (on the right, defined in eq.\ (\ref{sus}))
for $N=16,$ $R=4$ and $\lambda = 0.44$. The location of the
maximum of $C$ coincides with the peak in $\chi_{0}$, which provides
a consistent result for the critical value $m_{c}^{2} \simeq - 0.32\,$.}}
\label{chiC1}
\vspace*{-2mm}
\end{figure}

Let us describe the determination of $m_{c}^{2}$.
As a first example, Figure \ref{chiC1} shows results
at $N=16$, $R=4$, $\lambda =0.44$. The specific heat takes 
its maximum at $m_{c}^{2} \simeq - 0.32\,$, and the susceptibilities
confirm this value. The peak for $\chi_{0}$ further specifies
that we enter the uniform ordered phase for $m^{2} < m_{c}^{2}$.
Unlike $C$, the $\chi_{\ell}$ are sensitive to the type 
of order below $m_{c}^{2}$.

\begin{figure}[h!]
\begin{center}
\includegraphics[angle=270,width=.6\linewidth]{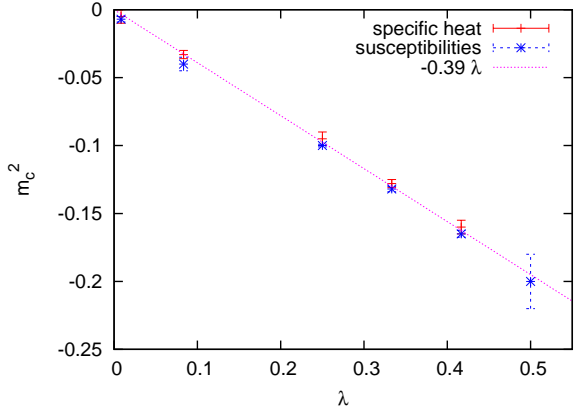}
\vspace*{-3mm} \\
\end{center}
\caption{{\it The order/disorder transition line in the phase diagram
for $N=12$, $R=8$. The transition is identified consistently
from two criteria. This figure shows the regime of weak coupling, 
where $m^{2} < m_{c}^{2}$ implies a uniform order.}}
\label{chiC2}
\vspace*{-2mm}
\end{figure}

This agreement between the two criteria gives a reliable determination
of $m_{c}^{2}$. Figure \ref{chiC2} shows this consistency
in the case $N=12$, $R=8$ for a variety of $\lambda$ values.
It also gives an overview of the phase diagram: as $\lambda$ rises,
$m_{c}^{2}$ moves to more negative values.
This relation is linear to a good approximation, as we are going
to discuss in Section 4.

For fixed values of $N$ and $R\,$, the $(\lambda , m^{2})$-plane contains
a triple point, which we denote as $(\lambda_{T}, m_{T}^{2})$.
It separates the regimes of weak coupling, $\lambda < \lambda_{T}$,
and of moderate or strong coupling, $\lambda \gsim \lambda_{T}$
or $\lambda \gg \lambda_{T}$.

\begin{figure}[h!]
\begin{center}
\includegraphics[angle=270,width=.49\linewidth]{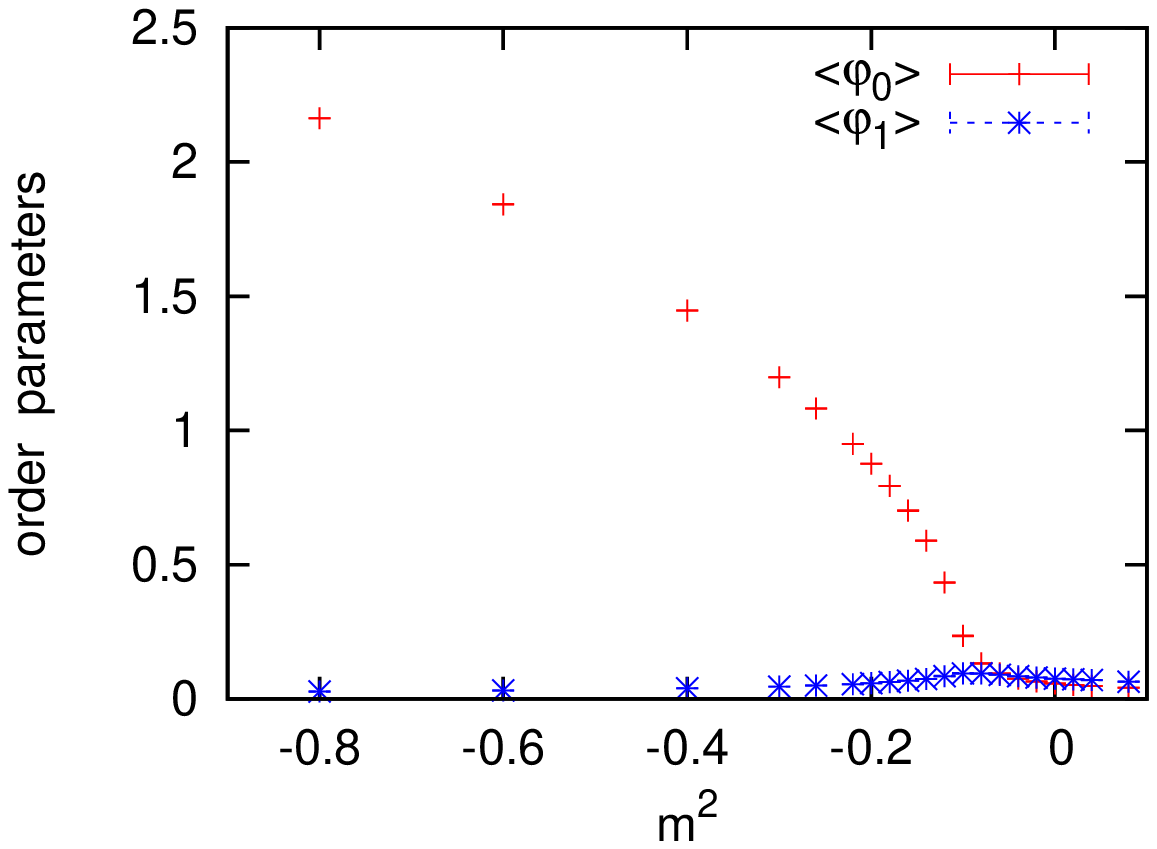}
\includegraphics[angle=270,width=.49\linewidth]{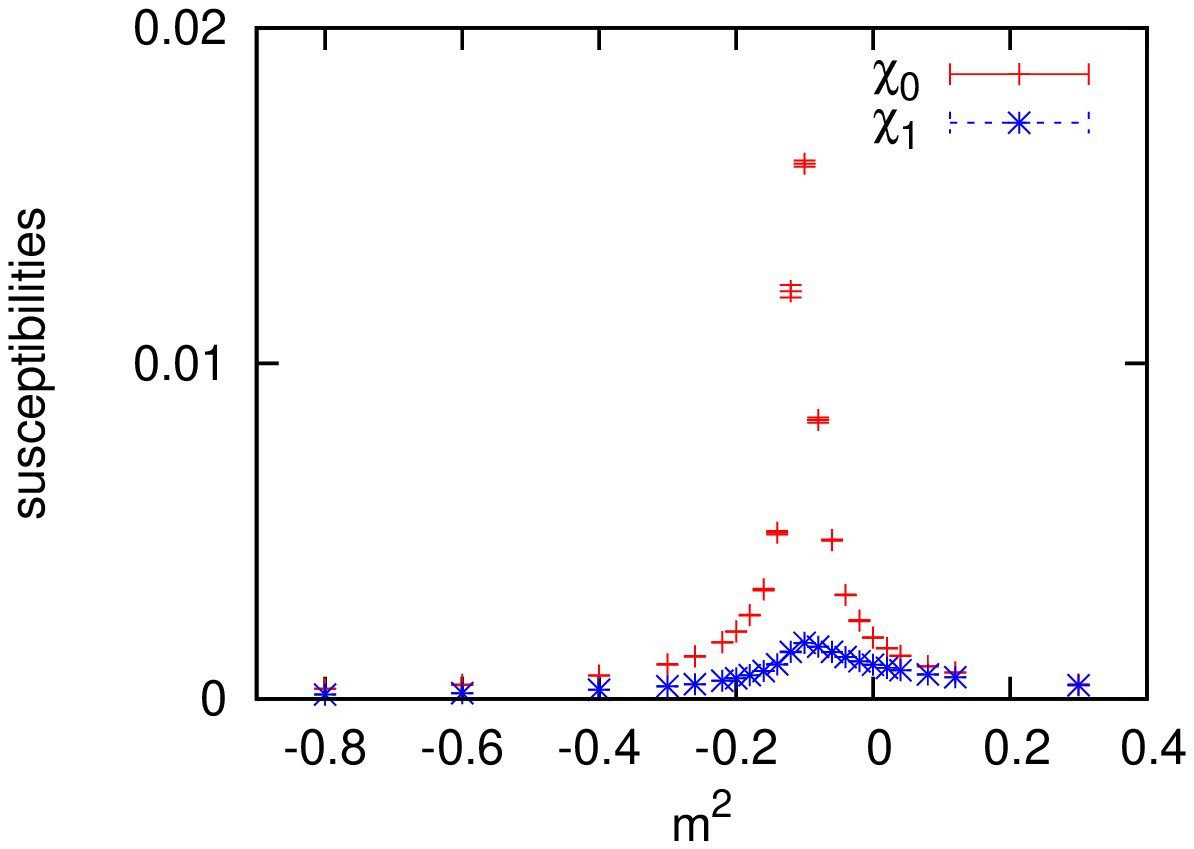}
\\
\includegraphics[angle=270,width=.5\linewidth]{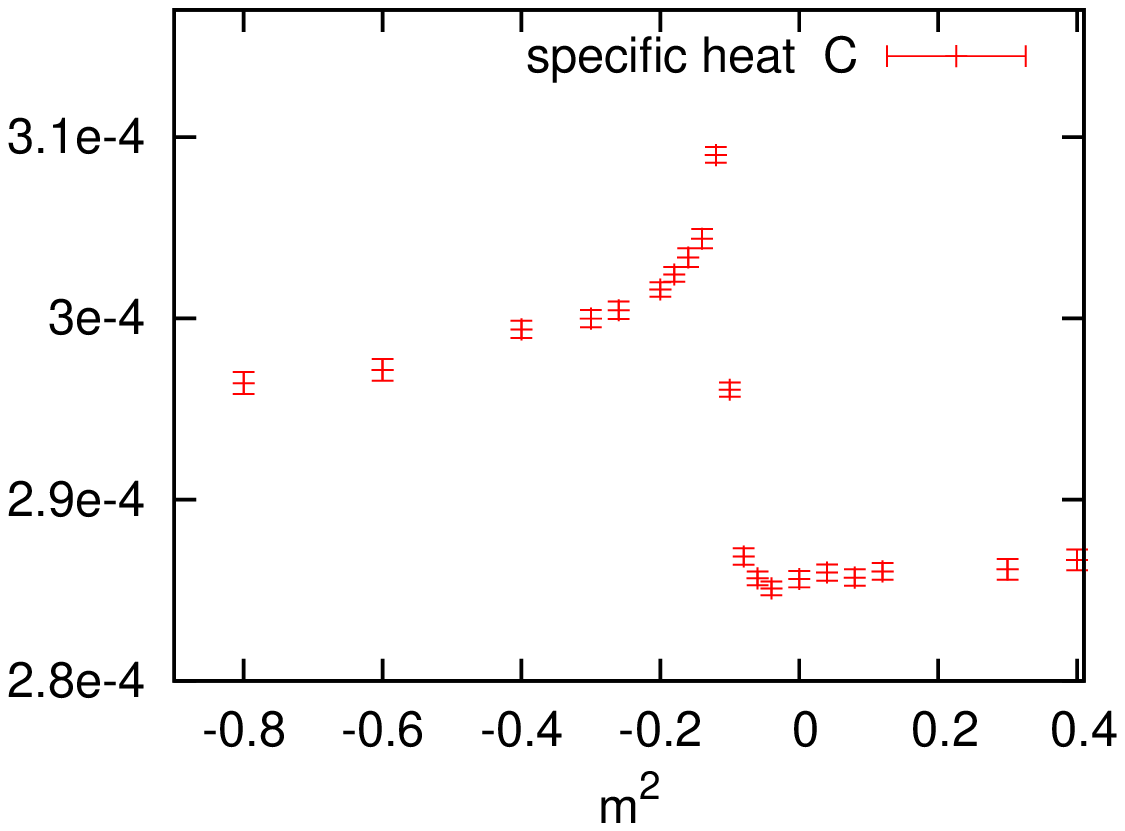}
\includegraphics[angle=270,width=.49\linewidth]{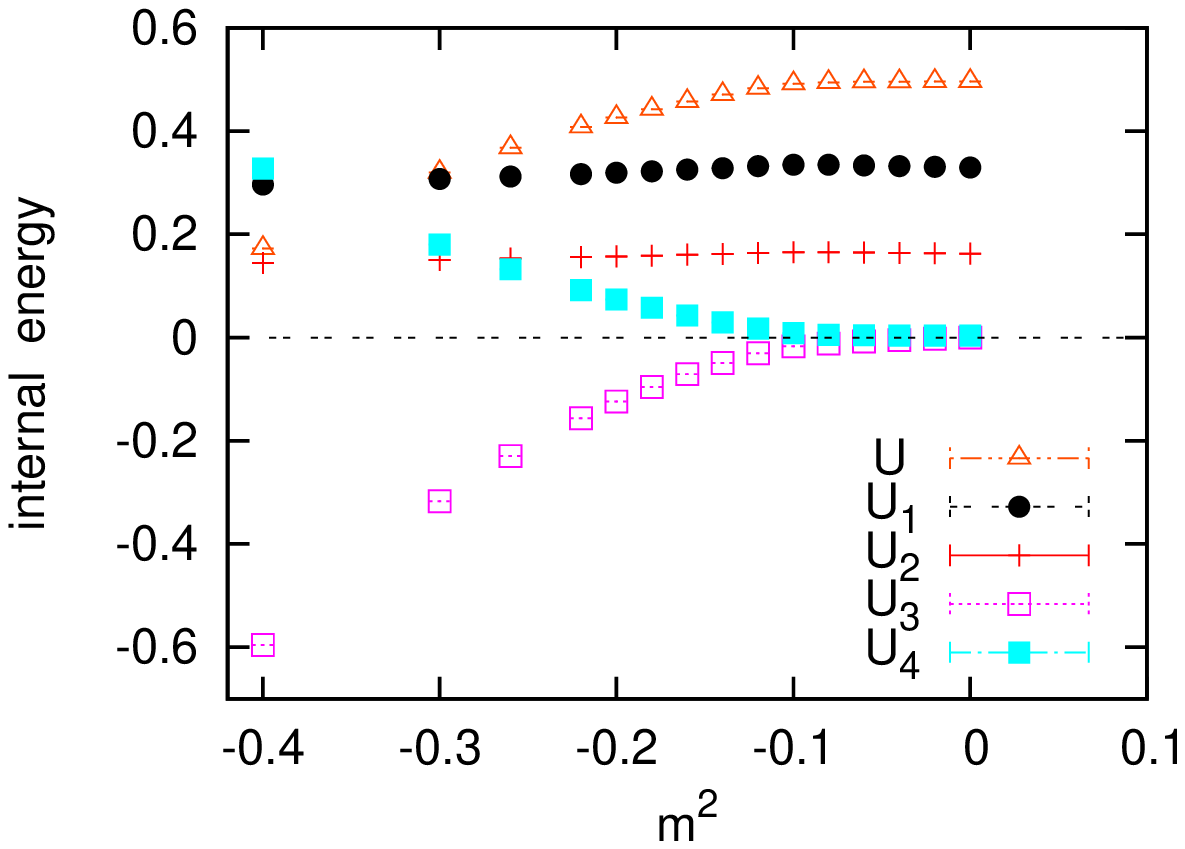}
\vspace*{-4mm} \\
\end{center}
\caption{{\it Simulation results for the determination of $m_{c}^{2}$ 
at $N=12$, $R=4$, $\lambda =0.17$. 
On top we show the order parameters according to 
specification (\ref{ordpar}) (on the left) and the corresponding 
susceptibilities (on the right).
We observe a transition between disorder and uniform order at
$m_{c}^{2} = -0.12(2)$. \newline
Below: the specific heat (on the left) and different contributions 
to the internal energy $U =   
\sum_{i=1}^{4} U_{i}$ (on the right). 
In the phase of uniform order, the potential terms deviate
significantly from zero, so that the mass term $U_{3}$
(the quartic term $U_{4}$) becomes negative (positive).
The kinetic contributions $U_{1}$ (spatial) and $U_{2}$ (temporal)
remain almost constant --- here each dimension contributes
approximately the same amount.}}
\label{weak}
\end{figure}

A typical example for the behaviour of the order parameters
at weak coupling is shown in Figure \ref{weak}. 
For sufficiently negative $m^{2}$ the order parameter 
$\la \varphi_{0} \ra $ rises drastically. The peaks in
$\chi_{0}$ and in the specific heat 
allow for a more accurate evaluation of $m_{c}^{2} = -0.12(2)$.
Further insight into this phase transition is gained by
splitting the internal energy $U$ (in eq.\ (\ref{sheat}))
into contributions due to the
different terms in the action (\ref{action}),
\begin{tabbing}
..................................................\= \kill
$U_{1} 
= \frac{2 \pi R^{2}}{N} \langle
{\rm Tr} [ \sum_{t} (\Phi_{t+1}-\Phi_{t})^{2} ] \rangle 
$ \> : spatial kinetic contribution \vspace*{5mm} \\
$U_{2}
= \frac{2 \pi }{N} \langle
{\rm Tr} [ \sum_{t} (\Phi_{t} \hat {\cal L}^{2} \Phi_{t})] \rangle 
$ \> : temporal kinetic contribution \vspace*{5mm} \\
$U_{3}
= \frac{2 \pi R^{2}m^{2}}{N} \langle
{\rm Tr} [ \sum_{t} \Phi_{t}^{2} ] \rangle 
$ \> : contribution due to the mass term \vspace*{5mm} \\
$U_{4}
= \frac{\pi R^{2} \lambda}{N} \langle
{\rm Tr} [ \sum_{t} \Phi_{t}^{4} ] \rangle 
$ \> : contribution due to the self-interaction.
\end{tabbing}
They fulfil the identities
\be
U_{1} + U_{2} + U_{3} + U_{4} = U \ , \quad
2 ( U_{1} + U_{2} + U_{3} + 2 U_{4}) = 1 \ ,
\ee
(the latter is obtained from a variational argument).
In our setting the total internal energy $U$ is identical to the 
entropy. The last plot in Figure \ref{weak} illustrates that
it deviates from a constant as $m^{2}$ is decreased 
below $m_{c}^{2}$.
Moreover we see that $U_{3}$ and $U_{4}$ drift away 
from zero at this transition:
the quadratic term  becomes negative and the quartic
term positive, which is just the situation that triggers 
spontaneous symmetry breaking with uniform ground states.

\begin{figure}[h!]
\begin{center}
\includegraphics[angle=270,width=.5\linewidth]{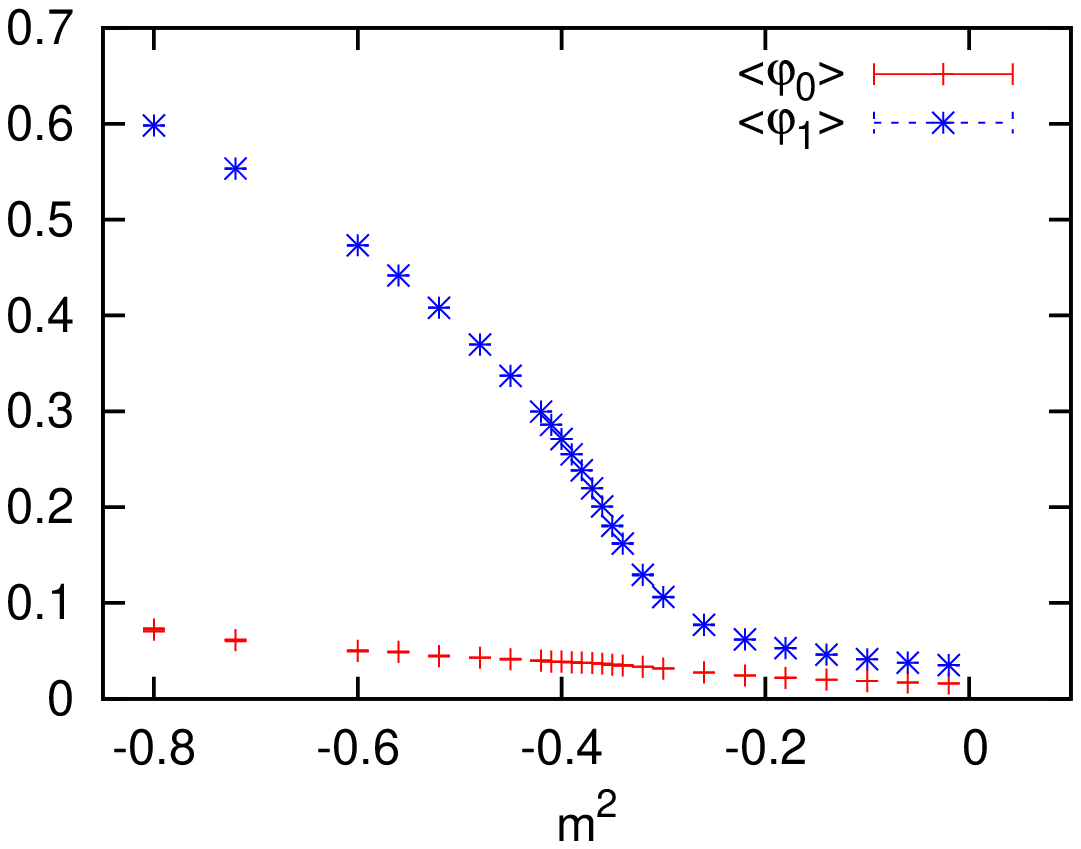}
\hspace*{-3mm}
\includegraphics[angle=270,width=.5\linewidth]{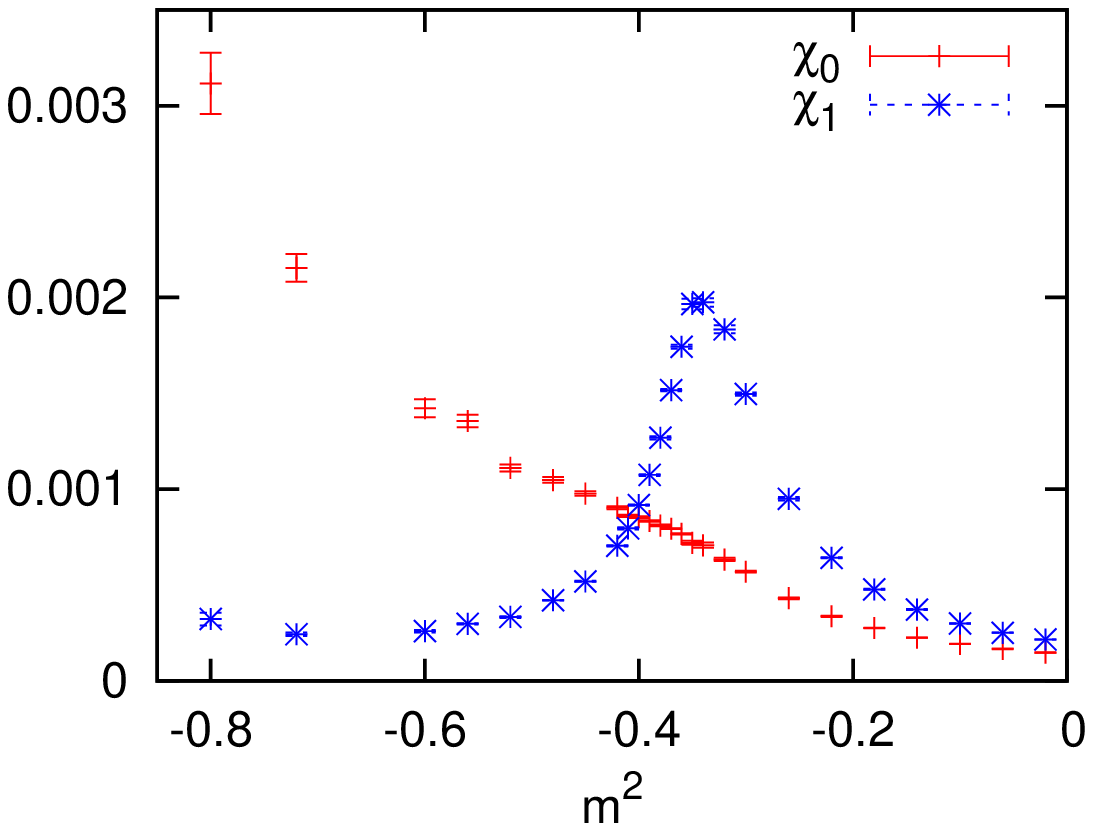}
\vspace*{-5mm} \\
\end{center}
\caption{{\it The order parameters and susceptibilities for an 
example at moderate coupling, $N=12$, $R=8$, $\lambda = 1.25$. 
At $m^{2} \simeq -0.37$ we observe a transition between
disorder and non-uniform order.}}
\label{strong}
\vspace*{-3mm}
\end{figure}

\begin{figure}[h!]
\begin{center}
\includegraphics[angle=270,width=.6\linewidth]{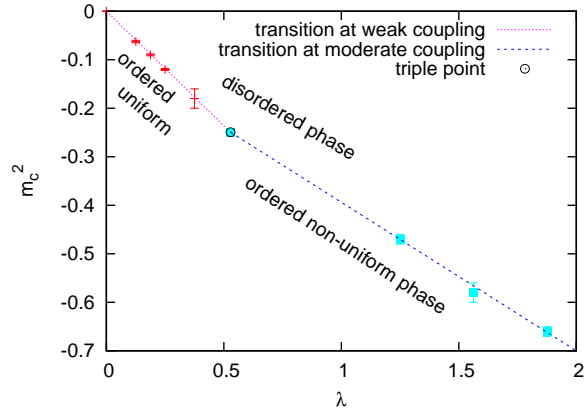}
\vspace*{-3mm} \\
\end{center}
\caption{{\it Overview of the phase diagram at $N=16$ and $R=8$.}}
\label{fulldia}
\end{figure}

Let us proceed to the regime of moderate coupling. Figure \ref{strong}
shows the order parameters for an example in that regime, along with
the susceptibilities. The data
reveal a transition between disorder and non-uniform order
at $m_{c}^{2} = -0.37(2)$. 
To provide an overview, we sketch in Figure \ref{fulldia} the 
complete phase diagram obtained at $N=16$, $R=8$, as a further example.

%% file: scaling.tex
In this section we analyse the scaling of the observed
phase transition lines --- and in particular their intersection
in the triple point --- with respect to $N$ and $R$.
We proceed by considering separately the boundaries 
of the disordered phase with the two ordered phases.

\subsection{Transition between the disordered and the
uniform ordered phase}

\begin{figure}[h!]
\vspace*{-2mm}
\begin{center}
\includegraphics[angle=270,width=.5\linewidth]{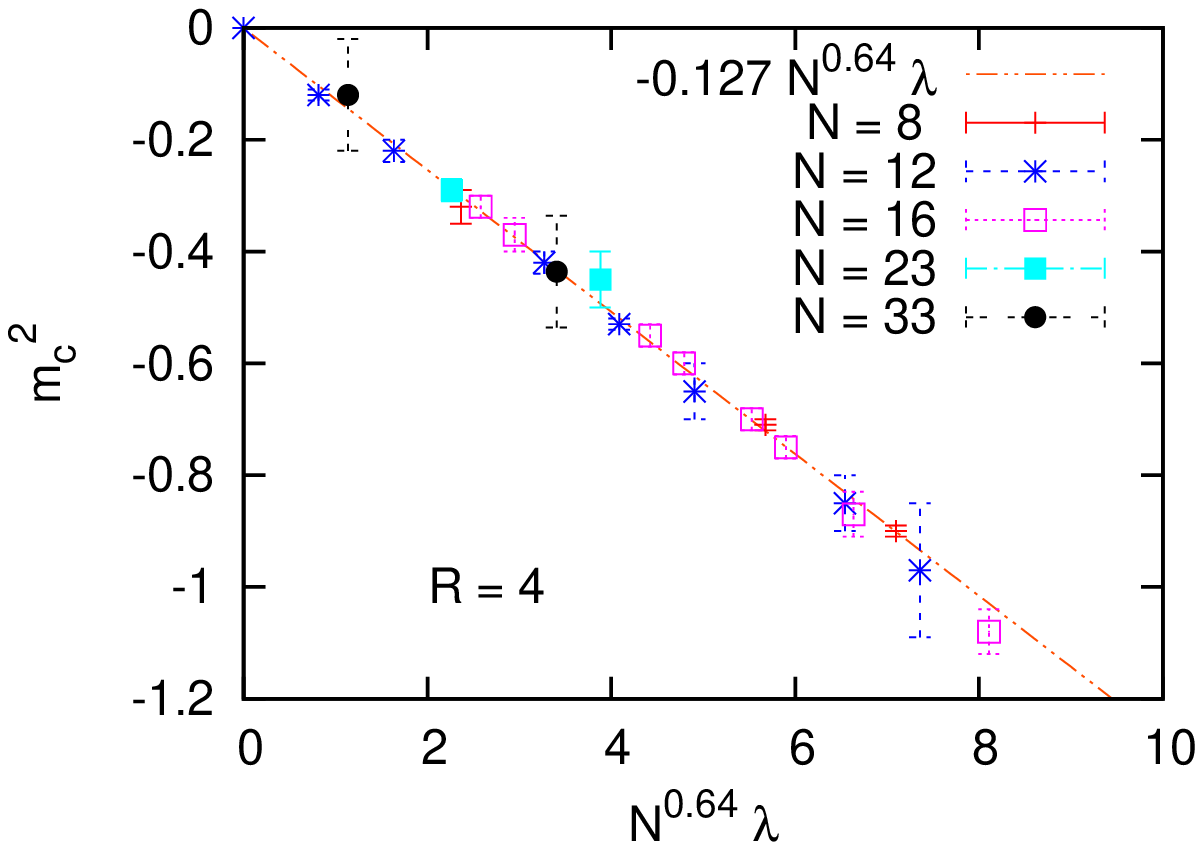}
\hspace*{-3mm}
\includegraphics[angle=270,width=.5\linewidth]{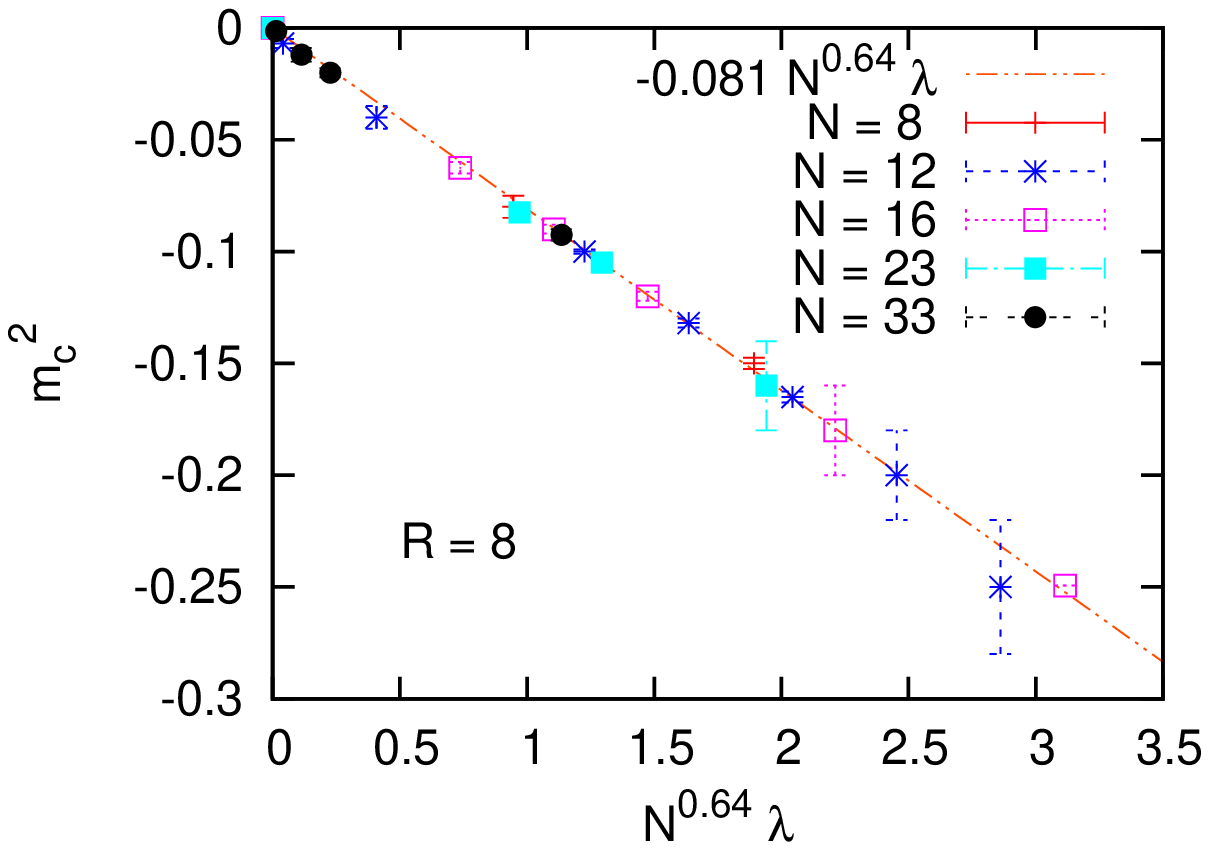}
\vspace*{-2mm} \\
\end{center}
\caption{{\it The phase transition disorder/uniform order 
at $N= 8 \dots 33$ for
$R=4$ (on the left), and for $R=8$ (on the right) .
We observe consistency with the exponent 
$\delta_{N} \simeq 0.64$ given in eq.\ (\ref{deltaN}).}}
\label{NR4-8dis-uni}
\end{figure}

\begin{figure}[h!]
\begin{center}
\includegraphics[angle=270,width=.5\linewidth]{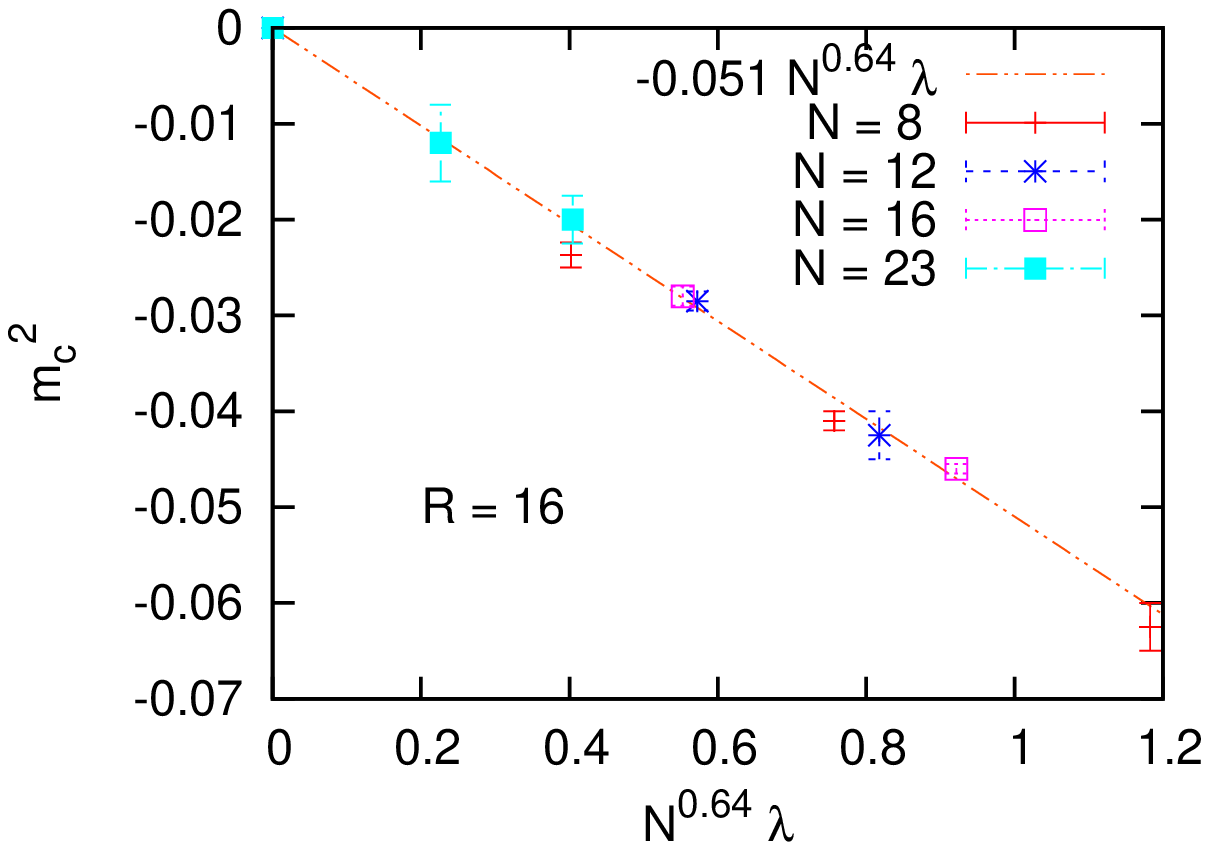}
\hspace*{-3mm}
\includegraphics[angle=270,width=.5\linewidth]{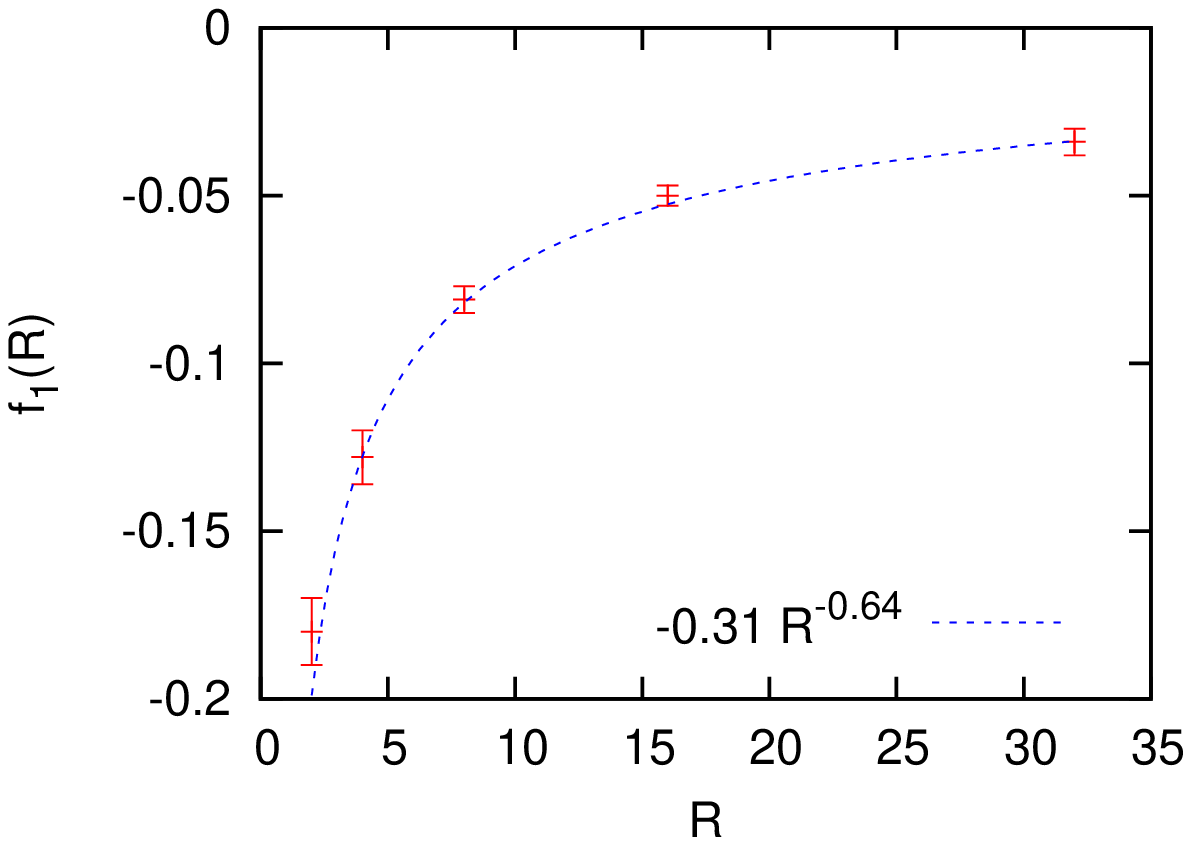}
\vspace*{-2mm} \\
\end{center}
\caption{{\it On the left: the transitions disorder/uniform order 
at $R = 16$ for a variety of $N$ values.
On the right: the fit for the function $f_{1}(R) := f(N,R)N^{-0.64}$,
which yields eq.\ (\ref{mc2disuni}) for the phase transitions
at weak coupling.}}
\label{NR16dis-uni}
\vspace*{-3mm}
\end{figure}

We first return to the fixed size $N=12$, $R=8$. Figure \ref{chiC2}
shows the measured disorder/uniform order transition line,
\be
m_{c}^{2} = -0.39(2) \lambda \ .
\ee
Probing also other sizes --- by varying $N$ and $R$ separately ---
we observed linear relations again, so we were guided to the ansatz
\be  \label{f1}
m_{c}^{2} = f (N,R) \, \lambda \ , \qquad
f (N,R) \propto N^{\delta_{N}} R^{\delta_{R}} \ .
\ee
In the last expression we anticipate that the function $f(N,R)$
can be parameterised successfully in a monomial form.
To illustrate this, we first fix again $R=8$ but vary $N$ in the range
$8 \dots 33$. Figures \ref{NR4-8dis-uni} and \ref{NR16dis-uni}
(on the left) show that we obtain
excellent fits with the exponent $\delta_{N} \simeq 0.64$
at $R=4 \dots 16$. By including similar
plots for $R=2$ and $32$ we extract
\be  \label{deltaN}
\delta_{N} = 0.64 (3) \ .
\ee
Subsequently we explore the dependence of $f (N,R)$ on $R$, and we
find again agreement with the monomial ansatz (\ref{f1}).
Moreover, it turns out that the function $f$
essentially just depends on the ratio $N/R$. The exponent 
$\delta_{R}$ and the coefficient are determined from the fit
in Figure \ref{NR16dis-uni} (on the right), and we arrive at
\be  \label{mc2disuni}
m_{c}^{2} = -0.31(1) \frac{N^{0.64(3)}}{R^{0.64(1)}} \, \lambda \ ,
\qquad ( \lambda \leq \lambda_{T} ) \ .
\ee
We add that the quality of the corresponding fits is very good
as long as $3/8 \lsim N/R \lsim 8$. If the ratio $N/R$ leaves this
interval on either side, the data request a more general function 
$f (N,R)$ --- beyond the monomial form --- though the ansatz for
$m_{c}^{2}$ as a linear function of $\lambda$ remains applicable over 
a wider range.

\subsection{Transition between the disordered and the
non-uniform ordered phase}

\begin{figure}[h!]
\vspace*{-4mm}
\begin{center}
\includegraphics[angle=270,width=.5\linewidth]{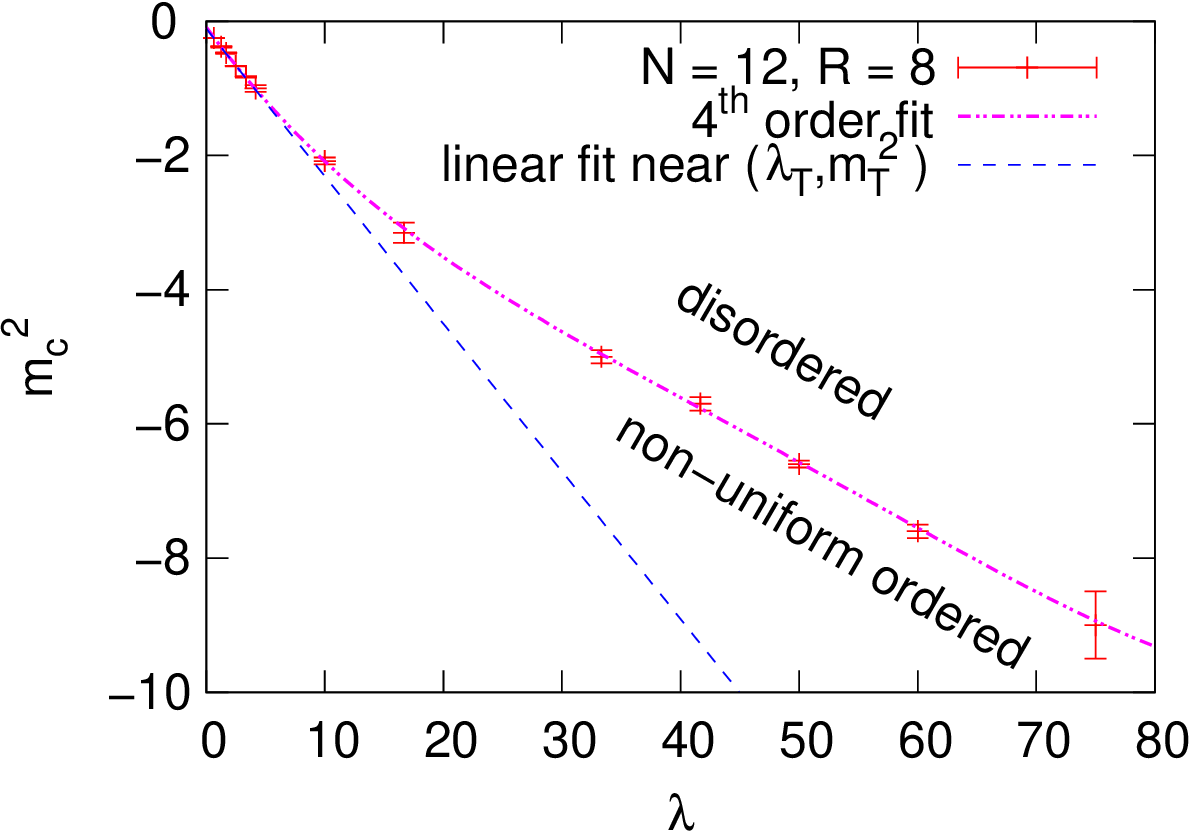}
\hspace*{-3mm}
\includegraphics[angle=270,width=.5\linewidth]{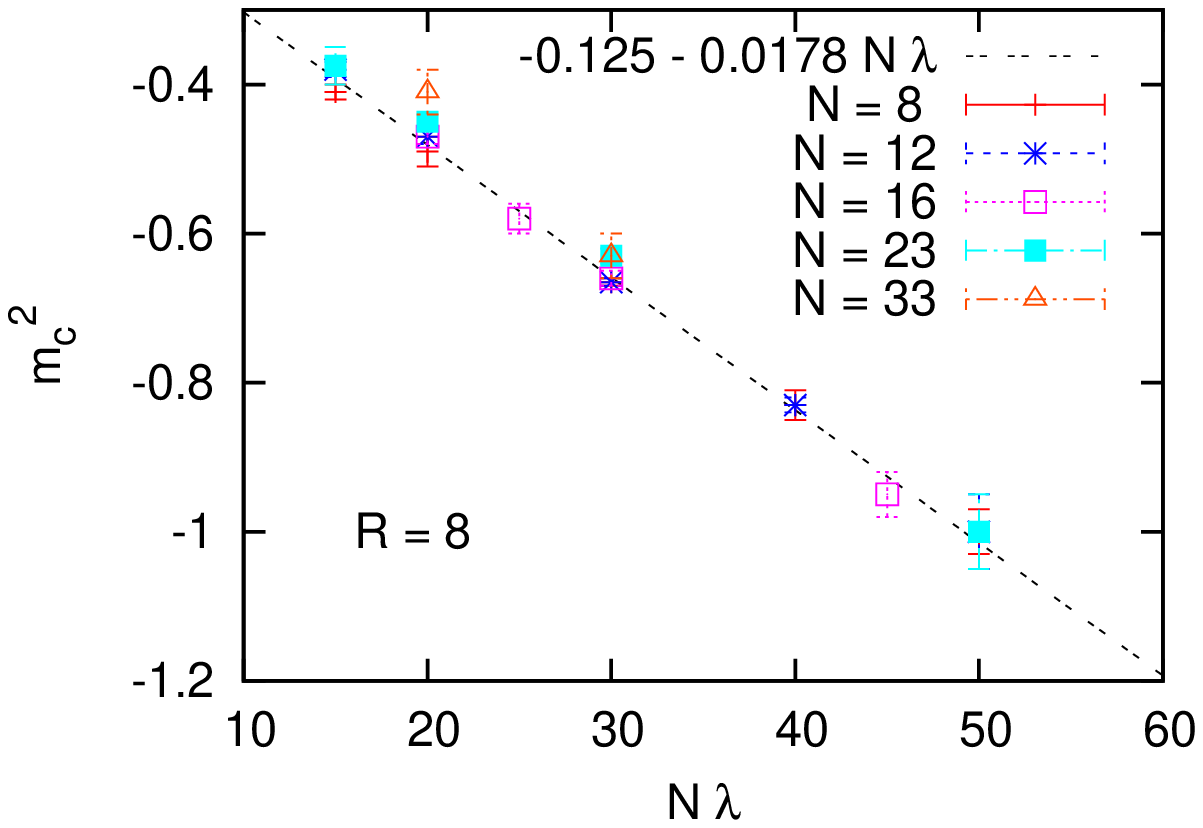}
\vspace*{-2mm} \\
\end{center}
\caption{{\it The transition disorder/non-uniform order 
at $R=8$ and $N=12$ (on the left), and $N= 8 \dots 33$ (on the right).}}
\label{NR8dis-nuni}
\end{figure}

Figure \ref{NR8dis-nuni} (on the left) shows the extension of Figure
\ref{chiC2} to much stronger couplings $\lambda$.
The critical values $m_{c}^{2}$ now mark the transition between
disorder and non-uniform order. Over this range of
$\lambda$ the phase transition line is curved.
However, at this stage we concentrate on the vicinity
of the triple point,\footnote{Large $\lambda$ values will 
be addressed in Subsection 5.3.} which is around
$(\lambda_{T}, m_{T}^{2}) \simeq (0.60 , \, -0.23)$ in this diagram. 
Up to $\lambda \approx 3$ a linear fit for $m_{c}^{2}$ is again
very precise,  but it now requires an additive constant,
$m_{c}^{2} = -0.11(2) - 0.22(1) \lambda\,$.

We extend also this consideration to $N = 8 \dots 33$, see
Figure \ref{NR8dis-nuni} on the right. A global fit implies
the generalised form
\be
m_{c}^{2} = -0.125(25) - 0.0178(10) N \lambda \ .
\ee

\begin{figure}[h!]
\vspace*{-4mm}
\begin{center}
\includegraphics[angle=270,width=.5\linewidth]{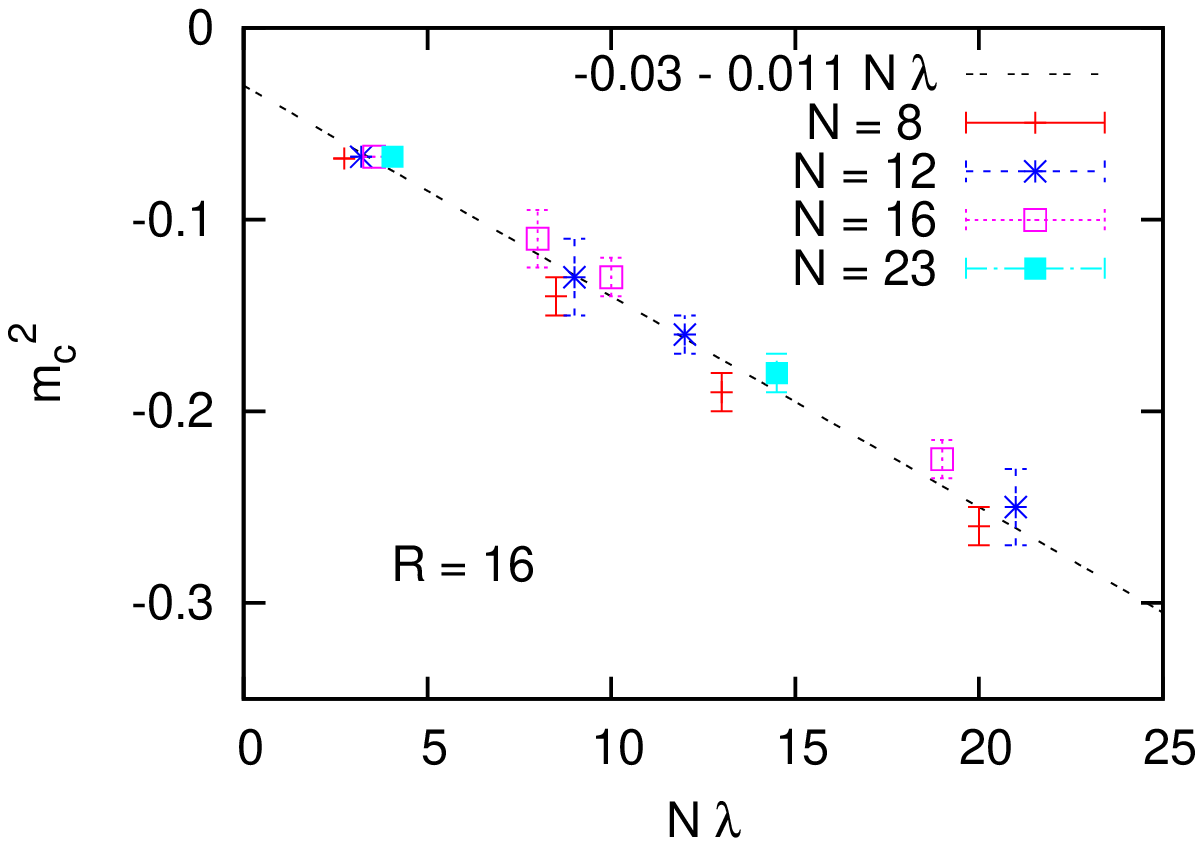}
\hspace*{-3mm}
\includegraphics[angle=270,width=.5\linewidth]{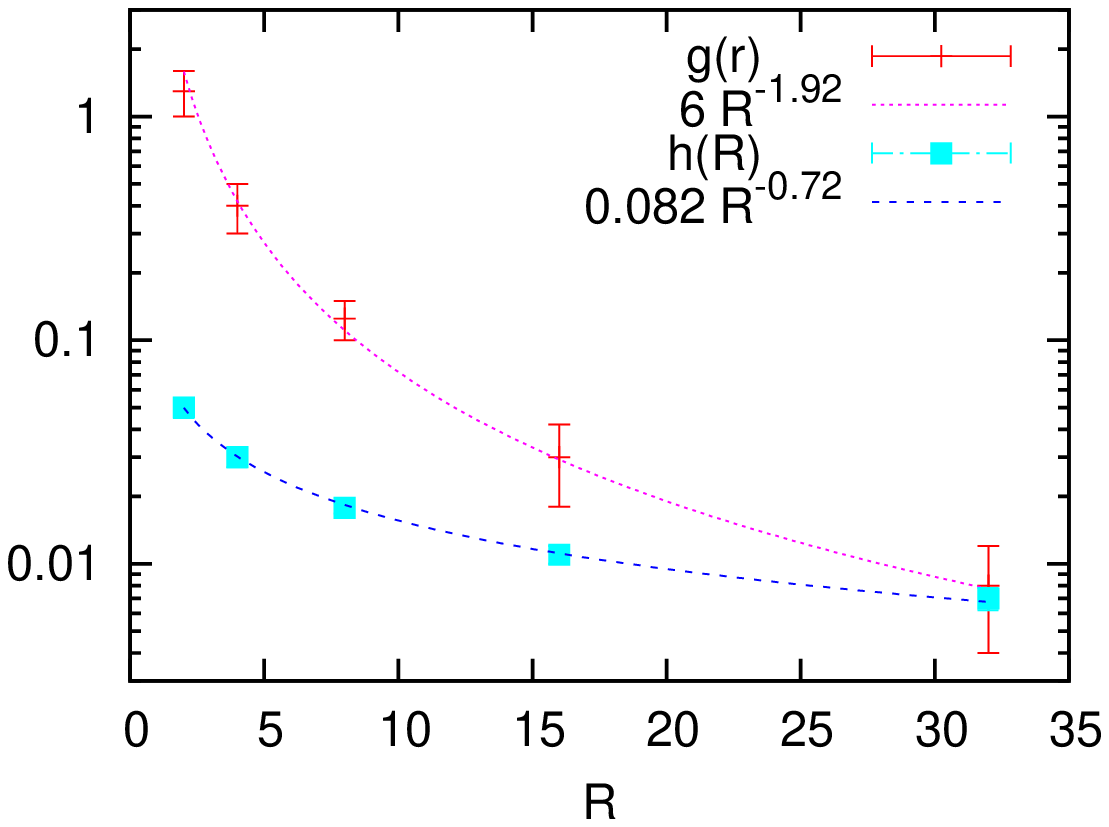}
\vspace*{-2mm} \\
\end{center}
\caption{{\it On the left: the transition disorder/non-uniform order 
in the phase diagram at $R=16$. On the right:
the fits for the functions $g(R)$ and $h(R)$, which lead to the
form (\ref{gheq}).}}
\label{NR16dis-nuni}
\end{figure}

In analogy to Subsection 4.1 we repeated this study for
$R=2,\, 4,\, 16$ and $32$; the case $R=16$ is shown as an example
in Figure \ref{NR16dis-nuni} (on the left). Since the choice of
$N \lambda$ as the parameter on the $x$-axis works well
in all these cases, we are led to the ansatz
\be \label{mc2disnuni}
m_{c}^{2} = -g(R) - h(R) N \lambda \ ,
\qquad ( \lambda \gsim \lambda_{T} ) \ .
\ee
Figure \ref{NR16dis-nuni} (on the right) shows
our fits for the functions $g(R)$ and $h(R)$, which imply
\be  \label{gheq}
g(R) = 6.0(15)R^{-1.92(10)} \ , \quad 
h(R) = 0.082(7) R^{-0.72(4)} \ .
\ee

\subsection{The triple point}

We are now in a position to evaluate the triple point
by the intersection of the phase
transition lines at low and moderate coupling, given in eq.\
(\ref{mc2disuni}), and in eqs.\ (\ref{mc2disnuni}), (\ref{gheq}). 
The intersection point is located at
\be  \label{tricri}
(\lambda_{T}, m_{T}^{2}) \simeq \frac{ \Big( 
19(5) R^{-1.20(11)} N^{-0.64(9)},
-5.9(16) R^{-0.56(11)} N^{0.0(1)} \Big) }
{R^{0.08(4)} - 0.26(2) N^{0.36(9)} } \ .
\ee
We assume Gaussian error propagation, and include a possible
error on the power of $N$ in the ansatz (\ref{mc2disnuni}).

This equation 
parameterises the triple point in the range for $N$ and $R$ that 
we simulated. However, this formula
cannot be extra\-polated to large $N$, due to the form of the 
denominator.\footnote{Unless one fixes $R \propto N$, such extrapolations
also takes us beyond the validity interval for the ratio
$N/R$ that we specified in Subsection 4.1.}
A direct investigation of the triple point --- and 
therefore of the phase diagram --- in the thermodynamic limit would 
require simulations at very large $N$.
We will come back to this issue in Section 6, where
we conjecture the properties of this extrapolation indirectly.

%% file: redu2.tex
Having elaborated the features of the phase diagram, we now compare
it to related models, which were also simulated 
in the recent years, and to reduced matrix models, which have been
studied analytically.

\subsection{The 2d $\lambda \phi^{4}$ model on a fuzzy sphere}

At finite $N$, the phase diagram for the 2d $\lambda \phi^{4}$ model
on a fuzzy sphere is qualitatively the same as we found in Section 3. 
An explicit convergence of our model to the 2d case could be expected in 
a setting which renders solely the temporal kinetic term negligible.
We saw, however, in Figure \ref{weak} that the impact of this term 
remains significant and approximately constant 
at weak and moderate coupling. This property is generic
in our study, hence there is no basis for expecting a reduction to the
2d model on a fuzzy sphere. 
At large $\lambda$ some reduction sets in, but it is of a different
kind, see Subsection 5.3.

\subsection{Comparison with a non-commutative torus}

Ref.\ \cite{BHN} presented a related numerical study
of the 3d $\lambda \phi^{4}$
model: also there the Euclidean time was lattice discretised,
while the spatial dimensions were treated by non-commutative 
coordinates. However, a lattice formulation and periodic boundary 
conditions were assumed for the spatial directions as well, and the
non-commutativity tensor $\Theta_{\mu \nu}$ was constant.
By means of Morita equivalence \cite{AMNS} the scalar field
--- first defined on a $N \times N$ lattice in the 
non-commutative plane --- 
was mapped onto Hermitian $N \times N$ matrices, so the
action ultimately simulated was similar to eq.\ (\ref{action})
(and also the rule (\ref{convent}) was the same).
The difference is the form of the spatial kinetic term: on the
non-commutative plane it was constructed by an adjoint matrix
operation which represents a shift by one lattice unit.

The phase diagram at finite $N$ was {\em qualitatively} equivalent to the
form that we found here. On the torus the non-uniform ordered
phase was denoted as ``striped phase'', and the stripe formation
for the two signs of $\phi_{t}(x_{1},x_{2})$ could indeed be visualised
by mapping the matrices back to 2d lattice configurations
(at a fixed time site).\\

Let us proceed to a {\em quantitative} comparison. 
On the torus the 
boundary of the disordered phase was approximately identified as \cite{BHN}
\bea
{\rm disorder/uniform~:} \quad &&
m_{c}^{2} \simeq -0.80 \lambda  \label{tor1} \qquad \qquad \qquad \qquad \\
{\rm disorder/non\!\!-\!\!uniform~:} \quad &&
m_{c}^{2} \simeq -0.48 \lambda - \frac{80}{N^{2}} \label{tor2} \\
{\rm triple~point~:} \quad && 
(N^{2} \lambda_{T} , N^{2} m_{T}^{2}) \approx (250 , \, -200) \ . \qquad
\label{tor3}
\eea
The transition disorder/non-uniform order was only explored
up to moderate coupling, $\lambda = O(1)$.

Our corresponding results are given in eqs.\ (\ref{mc2disuni}) and
(\ref{mc2disnuni}) to (\ref{tricri}).
The transition line (\ref{tor1}) can be matched if we set
$R \simeq 0.23 N$, but the triple point still differs
strongly. However, in the weak coupling region a link between
the two models can hardly be expected, due to the significance of 
the spatial kinetic term. 

We proceed to moderate coupling and look at the
requirements for agreement with eq.\ (\ref{tor2}).
The condition for the function $g(N,R)$ reads
$R \simeq 0.26 N^{1.04}$, which is close to
the above requirement for eq.\ (\ref{tor1}).
On the other hand, the condition due to
the function $h(N,R)$ deviates more from this
pattern, in particular in view of the exponent
($R \simeq 0.28 N^{1.39}$).

We anticipate at this point that the geometrical picture 
--- to be described in Section 6 --- suggests that the
non-commutative plane emerges for
\be
R \propto N^{\beta} \ , \quad \beta = 1/2 \ . \label{defbeta}
\ee
Hence the value for $\beta$ obtained at moderate $\lambda$
takes us even further away from the geometrical picture.
This may appear somewhat surprising, but it
is not paradoxical. For $\lambda/\lambda_{T} = O(1)$
the spatial kinetic term is significant, as the example in Figure
\ref{weak} shows, so a difference in this term may well
displace the phase transition lines.

In addition we saw that our results in Section 4 suffer from
finite $N$ artifacts, hence the considerations in this and the 
previous Subsection have to be interpreted cautiously.

\subsection{Reduction to a matrix chain}

At strong coupling $\lambda$ the impact of the kinetic
terms is reduced. In the extreme case where they are
fully negligible, one could imagine a transition to
a 1-matrix model of a single Hermitian
random matrix $\Phi_{1}$ with the potential 
$\, 4 \pi R^{2} \cdot {\rm Tr} \, [ \frac{m^{2}}{2} \Phi_{1}^{2} + 
\frac{\lambda}{4} \Phi_{1}^{4}] \, $.
That model has been studied analytically at $m^{2} \geq 0$ \cite{1matrix}
and at $m^{2} < 0$ with the result \cite{Shima}
\be  \label{1mateq}
m_{c}^{2} = - \frac{N}{R} \sqrt{\frac{\lambda}{\pi}} \ .
\ee
Consistent numerical data have been reported
for the 2d $\lambda \phi^{4}$ model on a fuzzy sphere
\cite{2dfuzzy} and on a non-commutative plane \cite{Jan}.
In our simulations we explored values up to 
$\lambda = O(10^{2} \dots 10^{3})$,
but we could not observe the feature of 
eq.\ (\ref{1mateq}) \cite{Julieta}. 

\begin{figure}[h!]
\vspace*{-2mm}
\begin{center}
\includegraphics[angle=270,width=.6\linewidth]{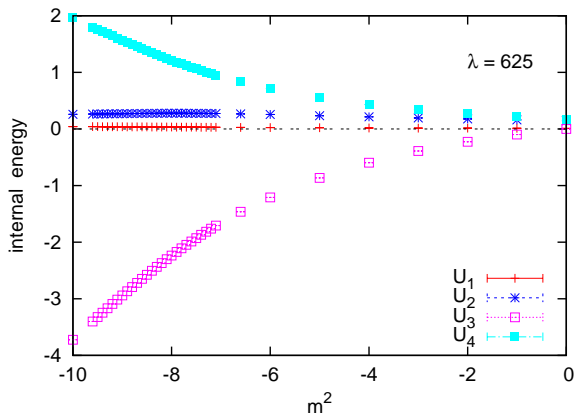}
\vspace*{-2mm} \\
\end{center}
\caption{{\it The different contributions to the internal energy at 
$N=8$, $R=16$ and $\lambda = 625$.
The temporal kinetic contribution $U_{2}$ now clearly dominates over
the spatial kinetic contribution $U_{1}$, in contrast to the
weak coupling behaviour shown in the last plot of Figure \ref{weak}.
This is fully consistent with the observed reduction
in the spatial directions only, which leads to the matrix chain
behaviour described in Ref.\ \cite{Shima}.}}
\label{strongU}
\end{figure}

We did find, however, consistency with a partial reduction,
which only neglects the {\em spatial} kinetic term; this is
opposite to the scenario commented on in Subsection 5.1.\footnote{One 
should expect the model on a non-commutative torus to agree in this regime, 
but it has only been simulated up to moderate coupling \cite{BHN}, 
as we mentioned in Subsection 5.2.}
In our formulation (\ref{action}) this means that the double-commutators 
$[{\cal L}_{i},[{\cal L}_{i},\Phi_{t}]]$ are negligible.
This behaviour is exactly confirmed by Figure \ref{strongU}.
Thus we obtain a {\em matrix chain} (analogous to a spin chain),
consisting of Hermitian matrices with a quartic potential, 
which are linked by a discrete second derivative. 
In the large $N$ limit Ref.\ \cite{Shima} derived the critical line
\be  \label{chaineq}
m_{c}^{2} = - \Big( \frac{3 N^{2}}{16 R^{2}} \, \lambda \Big)^{2/3} \ .
\ee
This formula corresponds to the one-cut vs.\ two-cut
transition in the eigenvalue distribution of
random matrices. This transition plays an essential r\^{o}le; 
it must show up at least for sufficiently large 
$\lambda$. In addition it is of importance
down to the triple point, where it merges
with the transition to the uniform order. 
But as the coupling is lowered, the fuzzy kinetic term 
distorts the form (\ref{chaineq}) for this transition line.
It eventually terminates at the triple point ---
a feature not present in the matrix chain.

\begin{figure}[h!]
\vspace*{-3mm}
\begin{center}
\includegraphics[angle=270,width=.5\linewidth]{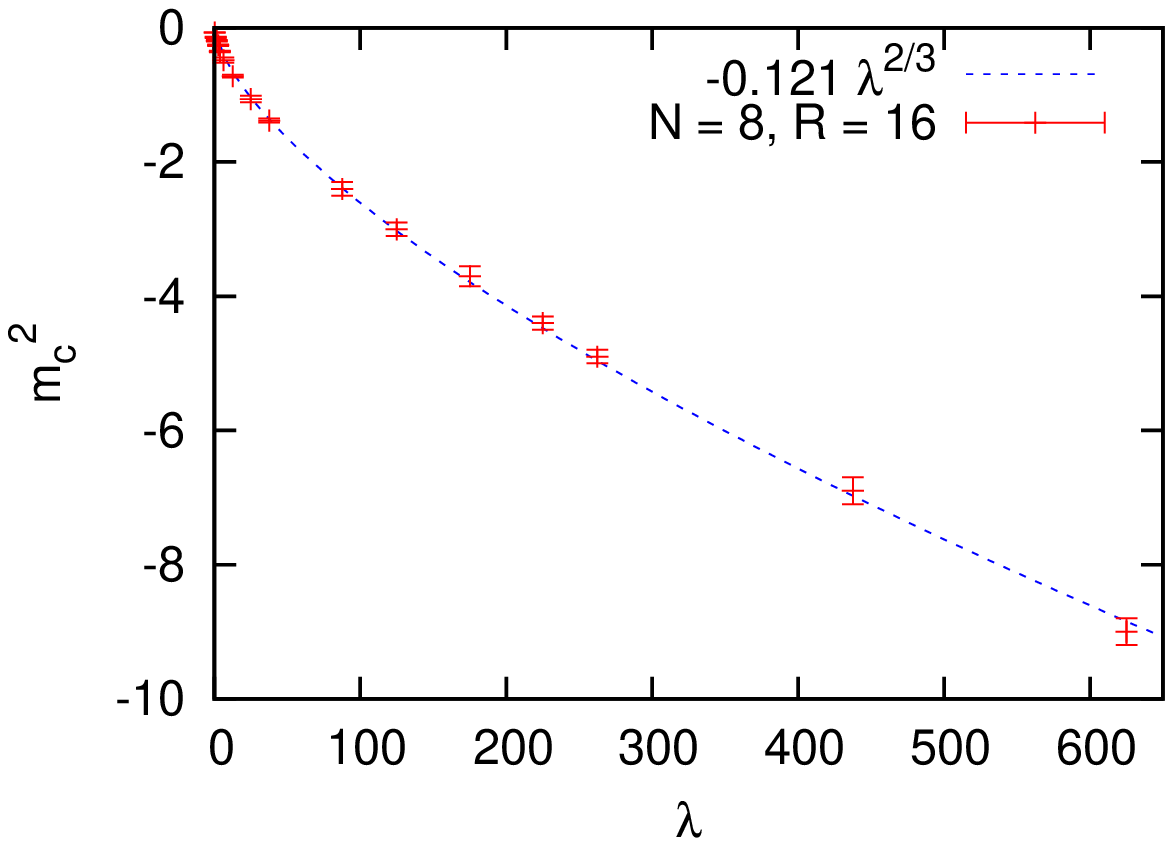}
\hspace*{-3mm}
\includegraphics[angle=270,width=.5\linewidth]{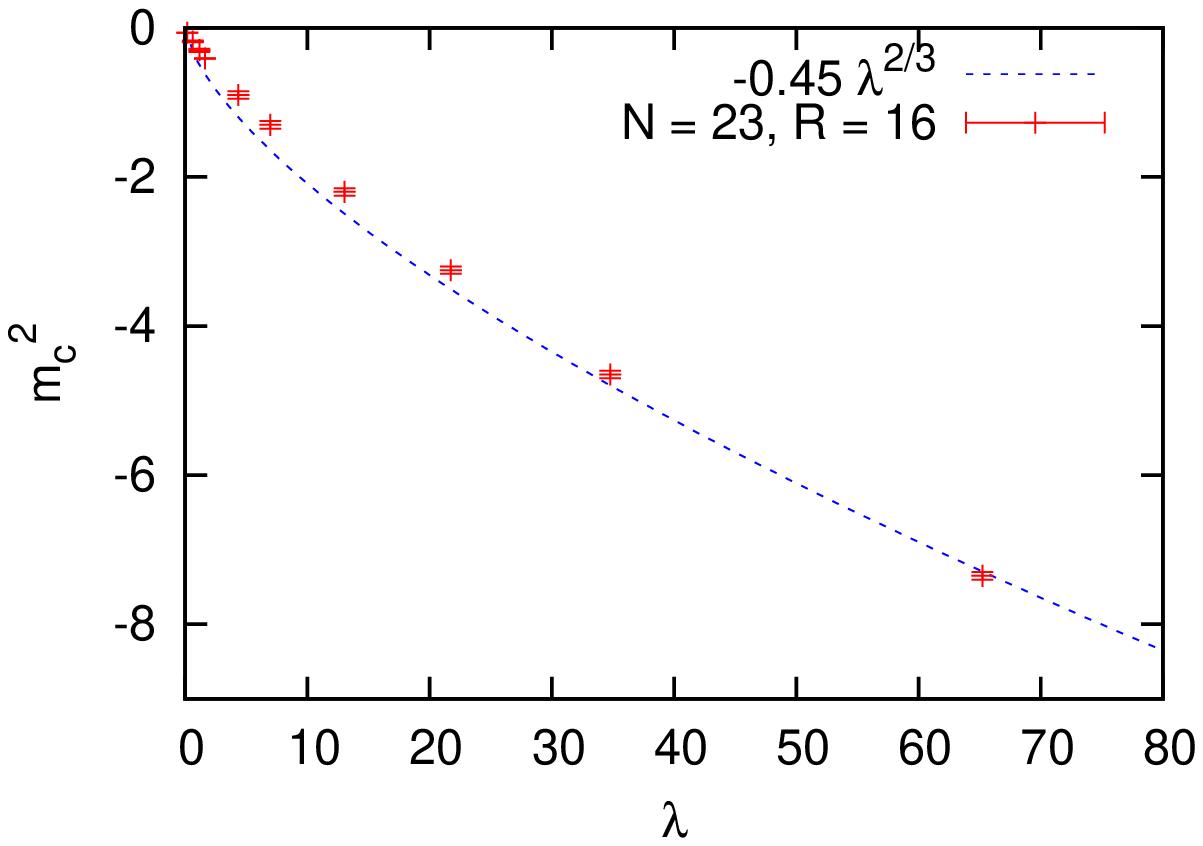}
\vspace*{-2mm} \\
\end{center}
\caption{{\it The phase transition at strong coupling $\lambda$ for
$N=8$, $R=16$ (on the left), and for $N=23$, $R=16$ (on the right).
In both cases we observe a broad window of agreement
with the matrix chain formula (\ref{chaineq}).}}
\label{strongplot1}
\end{figure}

\begin{figure}[h!]
\begin{center}
\includegraphics[angle=270,width=.53\linewidth]{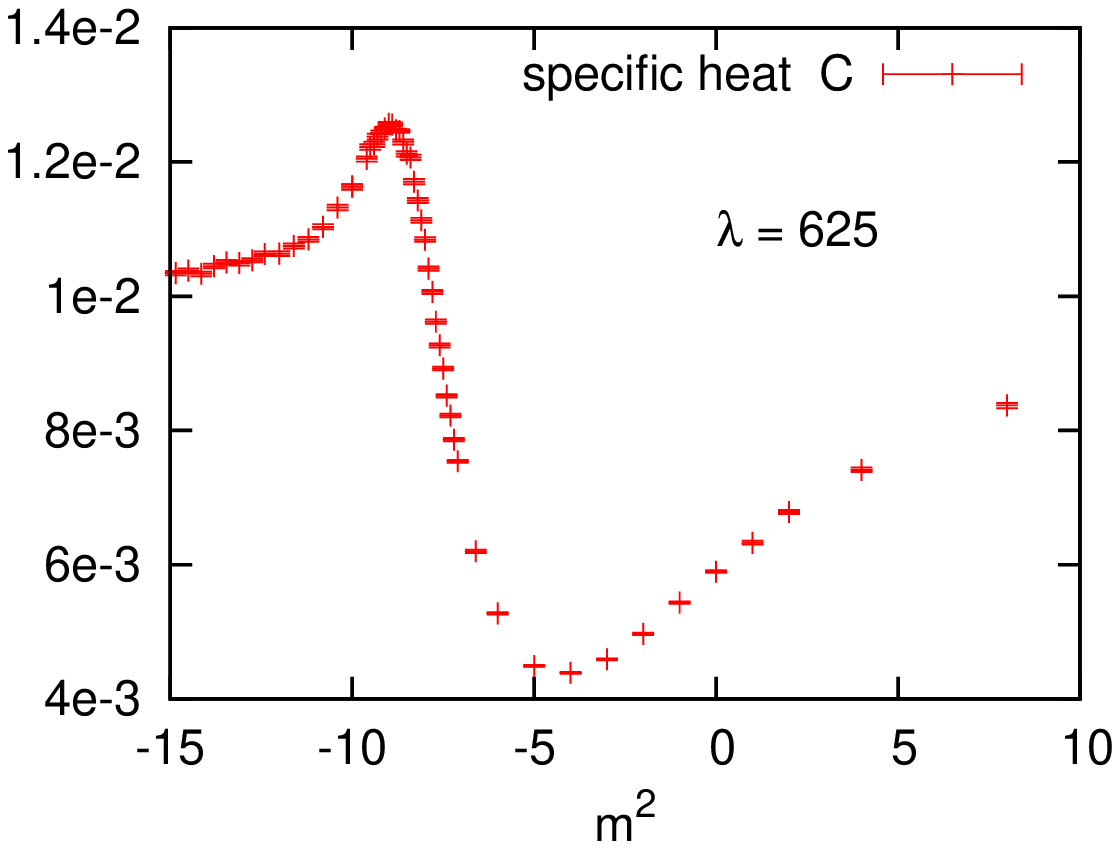}
\hspace*{-3mm} 
\includegraphics[angle=270,width=.47\linewidth]{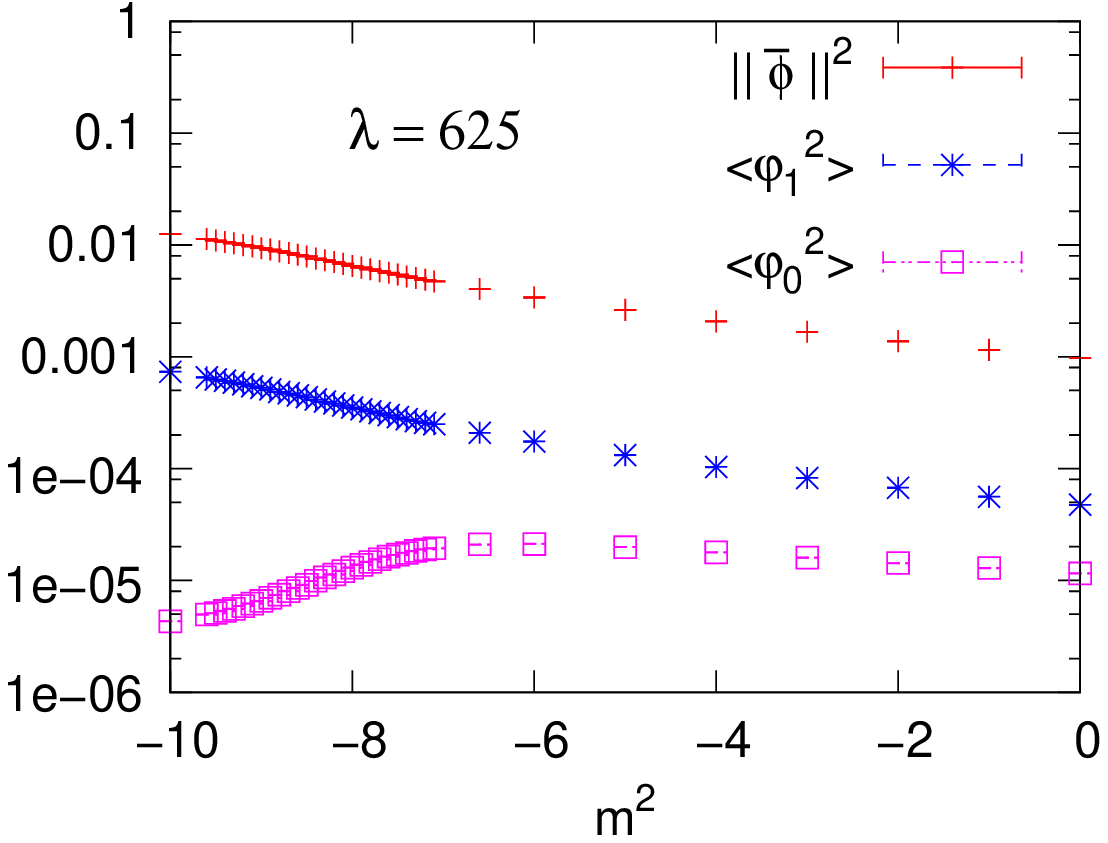}
\vspace*{-2mm} \\
\end{center}
\caption{{\it The specific heat (on the left) and the order parameters
(on the right) at $N=8$, $R=16$ and $\lambda = 625$.
We confirm the critical value $m_{c}^{2} = -9.0(5)$ at the
transition between disorder and a non-uniform order
(cf.\ Figure \ref{strongplot1} on the left). The plot on the
right demonstrates that it corresponds to a higher mode condensation, 
i.e.\ to a non-uniform order with a complicated pattern.}}
\label{strongplot2}
\end{figure}

Figure \ref{strongplot1} shows two examples where the
behaviour of eq.\ (\ref{chaineq}) is matched accurately,
up to a modification in the coefficient.
We take a closer look at the first case, $N=8$, $R=16$, where the 
precise agreement with the exponent in eq.\ (\ref{chaineq}) is
amazing because of the relatively small matrices.
We further fix $\lambda = 625$ and we show in Figure \ref{strongplot2}
the specific heat and the order parameters. The former confirms
the critical parameter $m_{c}^{2} = -9.0(5)$ which appears
in Figure \ref{strongplot1}. The order parameters 
demonstrate that we deal with a transition between disorder
and non-uniform order, where the latter
corresponds now to the condensation of higher modes, $\ell > 1$
(unlike the examples in Subsection 4.2).

This type of model (matrix quantum mechanics) has applications
in various branches of physics. For instance
QCD with $N_{f}$ light quarks flavours in a small box but an 
elongated time direction (the so-called ``$\delta$-regime'')
can be treated effectively by quantum mechanics 
of $SU(N_{f})$ matrices \cite{Heiri}. The very same technique 
can also be applied in solid state physics \cite{Has}.

Our case of Hermitian matrices attracted attention 
in string theory since the early nineties \cite{GroKle1,GroKle2}. 
In that framework it represents the $c=1$--model, which describes
random surfaces moving in one dimensions.
At a finite time periodicity $T$ there
is a vortices-driven Kosterlitz-Thouless phase transition
to the $c=0$--model \cite{GroKle1,c1c0}. Possible links to
QCD$_{2}$, to 2d black holes and to topological 
field theory were studied intensively \cite{GM}. 
This model continues to attract interest, 
see Refs.\ \cite{c1neu} for recent examples. 

An overview of the different reduction scenarios is added in Table
\ref{redutab}.

\begin{table}[h!]
\begin{center}
\begin{tabular}{|c|c||c|c|}
\hline
spatial  & temporal & setting &  status \\
kin.\ term & kin.\ term  & & \\
\hline
\hline
 & & 2d model on & numerical studies \cite{2dfuzzy}\\
large & small & a fuzzy sphere & not attained here \\
\hline
 & & matrix chain & analytical studies \cite{1matrix,Shima} \\
small & large & or c=1-model & reproduced here at large $\lambda$ \\
\hline
 & & total reduction & analytical study 
\cite{Shima} \\
small & small & 1-matrix model & not attained here \\
\hline
\end{tabular}
\end{center}
\caption{\it{An overview of the conceivable reductions
of our system --- due to the negligibility of kinetic terms ---
and their status in the literature and in this work.}}
\label{redutab}
\end{table}

%% file: conj.tex
In this section we discuss the extrapolation of the phase diagram
to large $N$ (which represents the thermodynamic limit) and to large $R$
(the transition of the spatial part to a plane).
In addition our rule (\ref{convent}) connects the large $N$
limit with the thermodynamic limit in the temporal direction.

We first consider the geometry of the sphere under these 
limits. As we remove the cutoff $N$, the coordinates (\ref{fuzzycor})
describe different 2d spaces, depending
on the simultaneous treatment of the radius $R$:

\begin{itemize}

\item The limit $N \to \infty$ at $R = const.$ leads to
a sharp sphere.

\item If the radius grows slowly, 
$R \propto N^{\beta}$, $0 < \beta < 1/2$, 
we end up with a sharp plane in the large $N$ limit.

\item If we take instead $N \propto R^{2} \to \infty$ we obtain
a non-commutative plane with a constant non-commutativity tensor
$\Theta_{\mu \nu} = {\rm i} \, \theta \, \epsilon_{\mu \nu}$ 
(where $\mu , \, \nu \in \{ 1, 2 \}$). 
This can be seen for instance in the plane which emerges
around the point $(0,0,1)$,
\be
[ X_{1}, X_{2} ] \simeq {\rm i} \frac{2R^{2}}
{\sqrt{N^{2}-1}} \, \frac{X_{3}}{R} \quad
\Rightarrow \quad \theta = \frac{2 R^{2}}{N} \ .
\ee

\end{itemize}

The scaling behaviour of a field theory on this space,
and in particular its phase diagram, still has to
be investigated. For the parameter range simulated, this
was carried out in Section 4.
We now address the issue of the large $N$
extrapolation, which was postponed in Section 4.
For the disorder/non-uniform order transition, 
we found agreement with the eqs.\ (\ref{mc2disnuni}) 
and (\ref{gheq}) at moderate $\lambda$, and with
eq.\ (\ref{chaineq}) at strong $\lambda$
(up to a modest modification of the coefficient).
In either case $\lambda$ only occurs
in a product with a positive power of $N$ ($1$ resp. $4/3$).
This is consistent with the suppression of field fluctuations
at large $N$ or large $\lambda$. 
Based on this property, we conjecture that exploring
the large $N$ behaviour could be equivalent to the case 
of large $\lambda$ at the values of $N$ in our study.

\begin{figure}[h!]
\vspace*{-3mm}
\begin{center}
\includegraphics[angle=270,width=.5\linewidth]{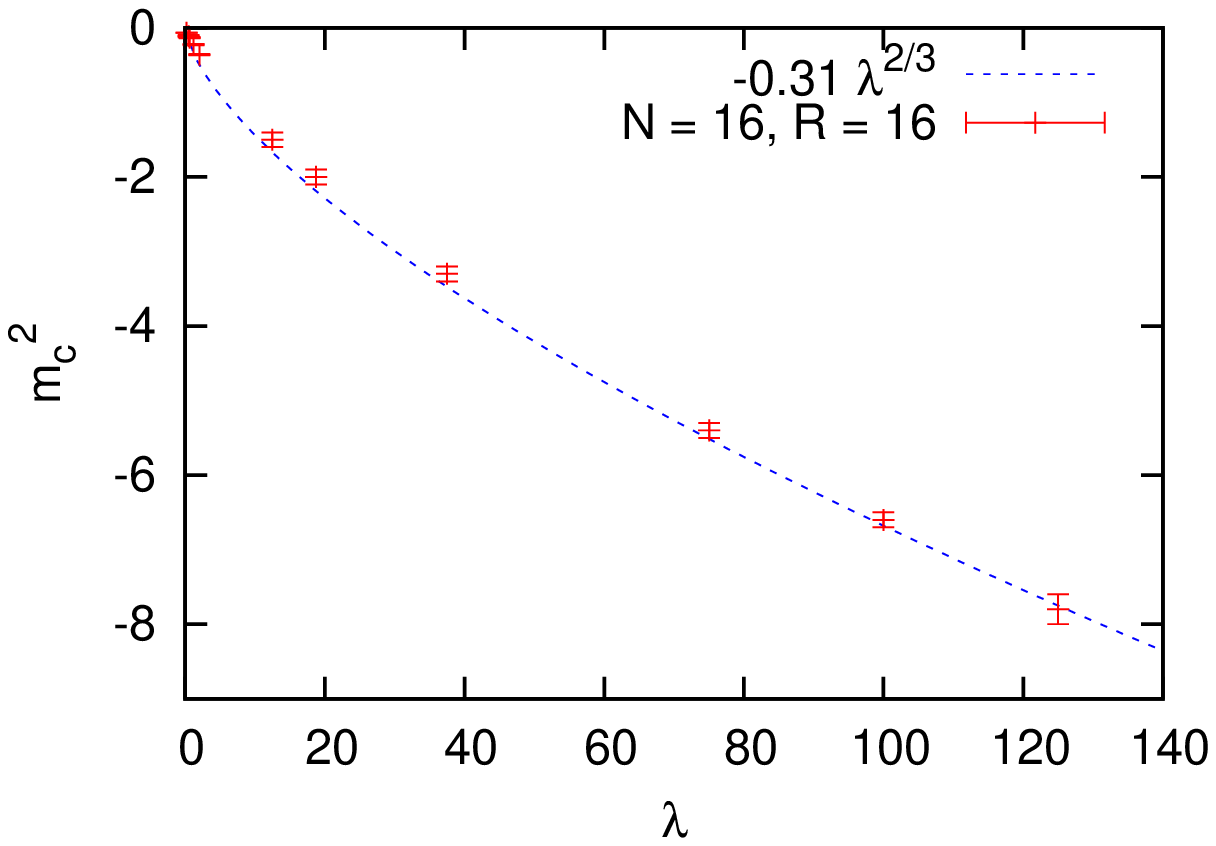}
\hspace*{-3mm}
\includegraphics[angle=270,width=.5\linewidth]{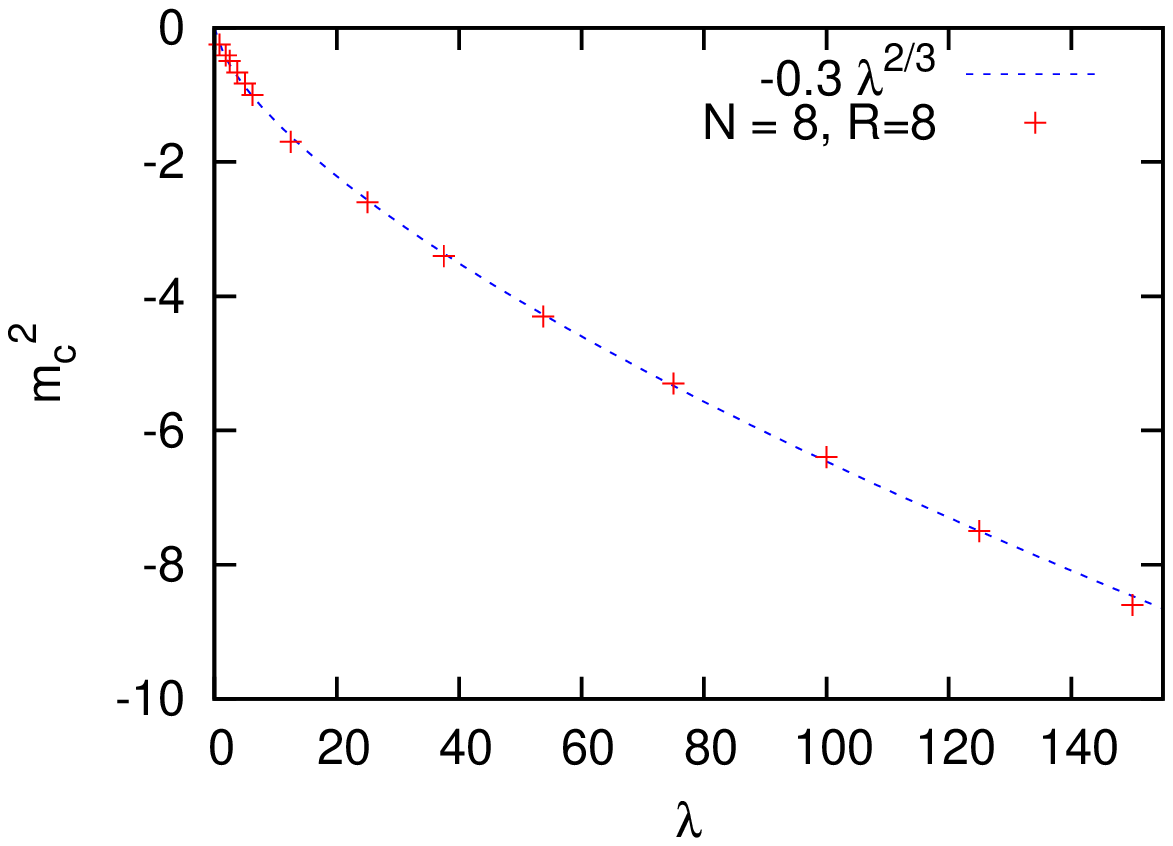}
\vspace*{-2mm} \\
\end{center}
\caption{{\it The phase transition at strong coupling $\lambda$ for
$N=16$, $R=16$ (on the left), and for $N=8$, $R=8$ (on the right).
As in Figure \ref{strongplot1} we find agreement 
with the matrix chain prediction
$m_{c}^{2} \propto \lambda^{2/3}$. In addition, comparison to
the result for $N=8$, $R=16$ in Figure \ref{strongplot1} 
supports the full proportionality relation (\ref{shima}).}}
\label{strongplotnew}
\end{figure}

This means that we now refer to eq.\ (\ref{chaineq})
to determine the triple point. The behaviour 
\be  \label{shima}
m_{c}^{2} = - c (N/R)^{4/3} \lambda^{2/3} \qquad (c = const.) 
\ee
is compatible with our data. For instance
at $R=16$ we obtained $m_{c}^{2}\lambda^{-2/3} \simeq -0.121$ 
at $N=8$, and $-0.31$ at $N=16$ (see
Figures \ref{strongplot1} and \ref{strongplotnew}),
which matches very well the behaviour $m_{c}^{2} \propto N^{4/3}$.
Also the dependence of the radius follows
eq.\ (\ref{shima}): for example the results at $N=8$ and
$R=8$ vs.\ $R=16$ (in Figures \ref{strongplot1} 
and \ref{strongplotnew}) agree very well
with the relation $m_{c}^{2} \propto R^{-4/3}$.
Considering many fits of that kind \cite{Julieta} we found
that both exponents and the coefficients $c$ in eq.\ (\ref{shima})
fluctuate within about $10 \%$ around the theortical
values of eq.\ (\ref{chaineq}), which were derived for the
matrix chain at large $N$ \cite{Shima}.

The intersection of this curve with the disorder/uniform order 
transition line (\ref{mc2disuni}) yields
\be  \label{lambdaT}
\lambda_{T} \propto \frac{N^{2.1(2)}}{R^{2.1(1)}} \ . 
\ee

In the framework of this conjecture, we obtain the
following scenarios\footnote{Based on the errors of the
exponents in eq.\ (\ref{lambdaT}), the distinction between
the scenarios should actually refer to $\beta = 1.0(1)$
(where $\beta$ is defined in eq.\ (\ref{defbeta})), but
for simplicity we just refer to $\beta =1$.}
for the limit $N \to \infty$:
\begin{itemize}

\item Limits with $N\to \infty$, $N/R \to \infty$
remove the phase of non-uniform order. Then the ordered
regime only consists of the uniform phase, which is
separated by an Ising transition from the disordered
phase. This limit corresponds to a commutative model. 
It includes in particular the case of
a sharp sphere ($R = const.$). 
In contrast to the tree-level expectation, this class of 
limits also captures the case $N \propto R^{2}$, which 
geometrically leads to a non-commutative plane.

\item If the limits $N,\, R \to \infty$ are
taken such that the ratio $N/R$ remains finite,
the triple point stabilises and the phase diagram
keeps qualitatively the form that we observed
at finite $N$; all three phases persist.

\item  Limits with $N\to \infty$, $N/R \to 0$
remove the phase of uniform order. Now the ordered
regime consists solely of the non-uniform phase.
This scenario is obtained for a rapidly expanding sphere.
Here the non-commutativity dominates the thermodynamic limit.

\end{itemize}

The transition lines that we identified at small and at moderate
coupling in Section 4 do not scale simultaneously for
for any fixed choice of the
axes of the phase diagram, as the forms (\ref{mc2disuni}) and
(\ref{mc2disnuni}) show. On the other hand, the large $N$ 
behaviour (as conjectured in this section) overcomes
this problem: if the triple point moves to $0$ or $\infty$,
only one transition line survives. In the case $R \propto N$,
which stabilises a finite triple point, the axes $(\lambda , m^{2})$
apply to both transition lines, without the necessity of rescaling.

%% file: conclu2.tex
We presented a numerical study of the phase diagram in the
3d $\lambda \phi^{4}$ model, where the spatial part is regularised
on a fuzzy sphere, while the Euclidean time is lattice discretised.
On the regularised level, we identified three phases based on the
order parameters $\varphi_{\ell}$ in eq.\ (\ref{phil}), 
the corresponding susceptibilities and the
specific heat. 
At fixed $\lambda$ there is a critical
parameter $m_{c}^{2} < 0$ : for $m^{2} > m_{c}^{2}$ 
($m^{2} < m_{c}^{2}$) the system is disordered (ordered).
The transition to $m^{2} < m_{c}^{2}$ leads to a uniform
order at small $\lambda$, and to a non-uniform order at moderate or
large $\lambda$.\footnote{We always refer to a region where $m^{2}$ is 
kept of the same magnitude as $m_{c}^{2}$; driving it to $m^{2} \ll
m_{c}^{2}$ causes simulation problems with the thermalisation
and decorrelation, hence we could not explore that region reliably. 
Similar technical problems obstructed a direct observation of 
the transition between the two ordered phases.} 

The boundary between these two scenarios corresponds to the 
triple point, which we denoted as $(\lambda_{T}, m_{T}^{2})$.
The transition disorder/uniform order (at $\lambda < \lambda_{T}$)
is analogous to a spontaneous magnetisation
and its critical line is parameterised by eq.\ (\ref{mc2disuni}). 

The non-uniform ordered phase emerges as a consequence of the 
non-locality in the fuzzy sphere regularisation.
That phase does not occur in a pure lattice regularisation. It corresponds
to a spontaneously broken rotation symmetry. At moderate coupling
strength, $\lambda \gsim \lambda_{T}$, the critical line is described
by eqs.\ (\ref{mc2disnuni}) and (\ref{gheq}). 
From its intersection with the curve
(\ref{mc2disuni}) we infer the location of the triple point
given in eq.\ (\ref{tricri}). This formula captures
a large amount of data that we collected, but it cannot be 
extrapolated to the limit $N \to \infty$.

Next we discussed the relation between the system studied
here and some models investigated in the literature, which have
qualitatively similar phase diagrams at finite $N$.
We found, however, significant differences from the 2d 
$\lambda \phi^{4}$ model an a fuzzy sphere \cite{2dfuzzy}. 
As for the 3d $\lambda \phi^{4}$ model on a 
non-commutative torus \cite{BHN} the vicinity of the triple point
can be matched rather roughly for a suitable relation 
between $N$ and $R$. 

At large $\lambda$, we observed consistency with the reduction to 
a matrix chain model, which had been solved in the large $N$ limit
\cite{Shima}. In particular our data match precisely the
predicted relation $m_{c}^{2} \propto (N^{2} \lambda /R^{2})^{2/3}$.
That model is known as the $c=1$--model in string
theory \cite{GroKle1,GroKle2,c1c0,GM,c1neu}, which we have 
therefore captured non-perturbatively. 

We then conjectured that large $N$ values may be equivalent
to large $\lambda$ couplings at moderate $N$, since $N$ and
$\lambda$ tend to appear only as products in the formulae for the
phase transition lines. 
This conjecture leads to different limits depending on the exponent 
$\beta$ in the relation $R \propto N^{\beta}$. For $\beta <1$ we obtain
a commutative limit, with an Ising-type transition between disorder
and uniform order. On the other hand, a rapidly expanding
sphere ($\beta >1$) leads to a dominance of the non-uniform phase.
That feature is a characteristic for a non-commutative theory,
where UV/IR mixing gives rise to an ordering due to the condensation
of a non-zero mode \cite{GuSo,CW,CZ,BHN}.

It remains an open question to verify the conjectured limits
by direct inspection of the triple point at very large system 
sizes $N$. For instance, at $R = const.$ one should verify 
the properties of the fuzzy sphere at larger $N$.
In the light of gravity-induced non-commutativity
on the Planck scale \cite{DFR}, a relation to quantum effects 
in a black hole might be conceivable
(see e.g.\ Refs.\ \cite{c1c0,black} for this line of thought).

According to our conjecture, the distinction 
between the scenarios of a commutative and a 
non-commutative limit is not located at the point where it 
is expected on purely geometrical grounds ($\beta = 1/2$).
Furthermore, it does not coincide with the picture suggested 
by a perturbative calculation to two loops \cite{fuzzyphi4theo1}. 
In that picture the commutative continuum limit could not be retrieved
at all because of the UV/IR mixing \cite{fuzzyphi4theo1}.\footnote{The 
model studied perturbatively in Refs.\ \cite{fuzzyphi4theo1}
coincides with the one considered here, up to the use of
a continuous Euclidean time. In the framework of the $c=1$--model
(at large $\lambda$)
it is a mystery that this difference seems to imply a $T \sim 1/T$
duality, which is absent in the matrix chain \cite{GroKle1}.}

A general lesson is that non-locality --- once it is introduced --- 
can cause surprises in the extrapolations
on the non-perturbative level.
This observation may serve as a warning also for the use of 
non-local actions in lattice simulations,
such as rooted staggered fermions (cf.\ footnote \ref{stagg})
or overlap fermions \cite{Neu} at strong gauge coupling.\footnote{Despite
the inverse square root in the overlap operator, it is local at
weak gauge coupling resp.\ on fine lattices \cite{HJL}. 
The range of locality --- and therefore of a safe definition of chiral 
fermions --- can still be enlarged by a non-standard kernel \cite{WB}.}
On the other hand, such surprises may
provide insight into other universality classes of interest, 
as we have seen. However, using the fuzzy sphere
as a regularisation scheme in quantum field theory is not
straightforward and its application
requires a careful investigation of the phase diagram.

%% file: simu2.tex
Our simulations were based on the Metropolis algorithm.
In each step we updated only one pair of conjugate matrix elements,
$(\Phi_{t})_{ij} = (\Phi_{t})_{ji}^{*}\,$. A detailed
comparison in technically similar simulations revealed that
this is far more efficient than updating complete matrices 
$\Phi_{t}$ \cite{BHN}. It proved useful to propose independent
changes of the real and the imaginary part of $(\Phi_{t})_{ij}$
with absolute values below $N \sqrt {m^{2}/\lambda}$ (and flat
probability distribution); this led to acceptance rates typically
around $1/2$. One sweep applies this step
successively to all independent elements in a configuration $\Phi$.

We were often confronted with several local minima
of the action. In many cases these minima were too pronounced for
a tunnelling to occur even in histories involving $O(10^{7})$
sweeps. This property obviously obstructs the direct measurement
of observables. As a first remedy we performed a number of runs
with independent hot starts. At the end we summed up the 
statistics collected in each run after thermalisation (which we 
are going to comment on below).
The histories in all runs had the same length of $O(10^{6})$ sweeps.
This improves the situation, but the number of runs was still 
too small (typically $O(10)$) to sample the vicinities of 
the different minima reliably. 

\begin{figure}[h!]
\vspace*{-3mm}
\begin{center}
\includegraphics[angle=270,width=.5\linewidth]{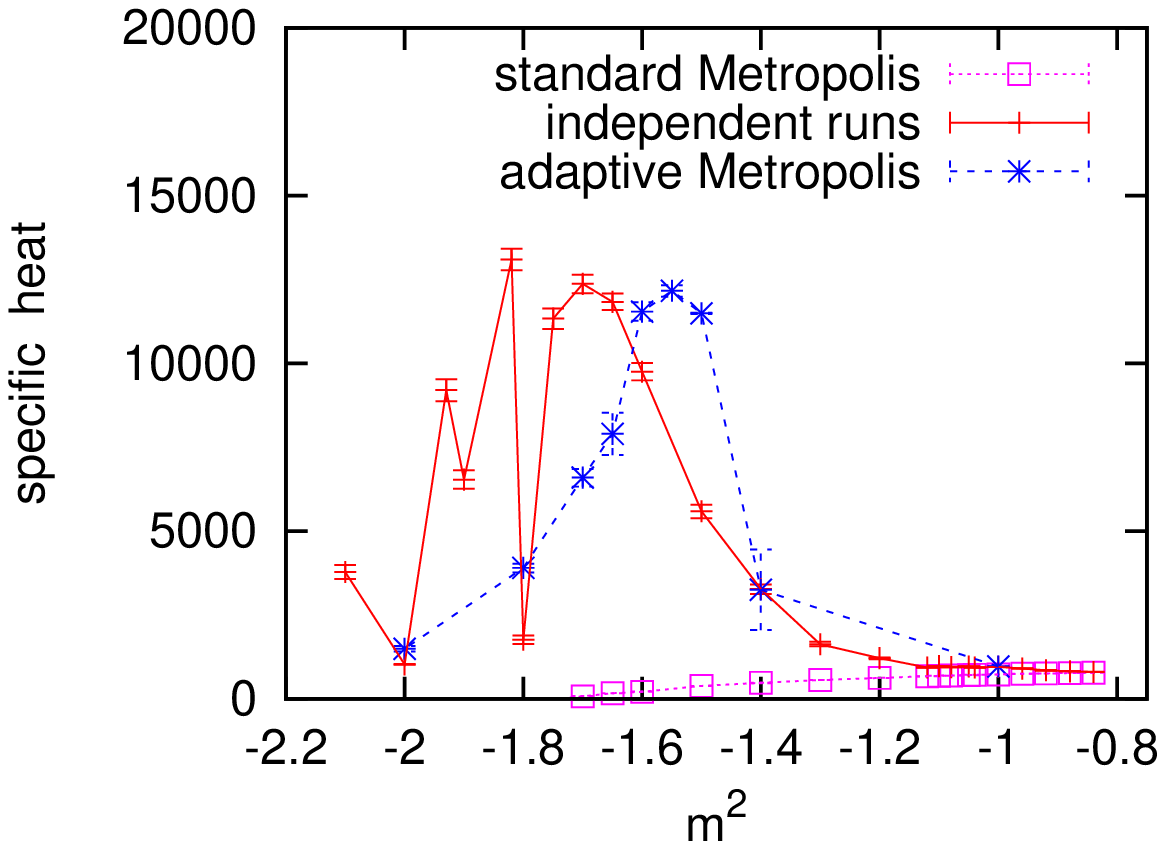}
\includegraphics[angle=270,width=.48\linewidth]{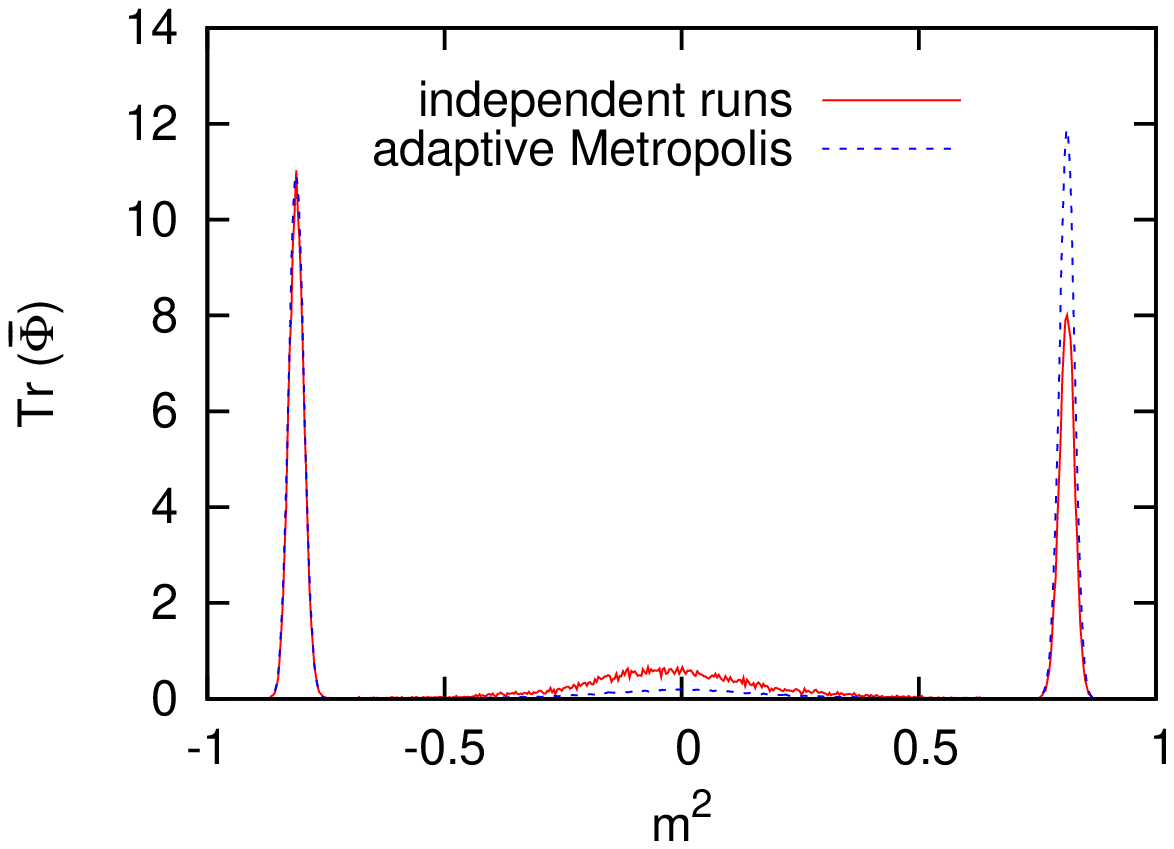}
\vspace*{-2mm} \\
\end{center}
\caption{{\it An illustration of the progress due to the adaptive
Metropolis algorithm. The parameters in these examples
are $N=12,$ $R=4$, $\lambda = 1.83$. On the left we show the specific
heat $C$; for adaptive Metropolis only the peak is clearly observed
at the critical value $m_{c}^{2}$. The plot on the
right is a histogram for ${\rm Tr} (\bar \Phi)$ at $m^{2} = -1.7$,
i.e.\ in the phase of uniform order. The two peaks approximate
the expected symmetric form well with adaptive Metropolis, but
not with the same number of independent runs.}}
\label{metro}
\vspace*{-3mm}
\end{figure}

\begin{figure}[h!]
\vspace*{-3mm}
\begin{center}
\includegraphics[angle=270,width=.49\linewidth]{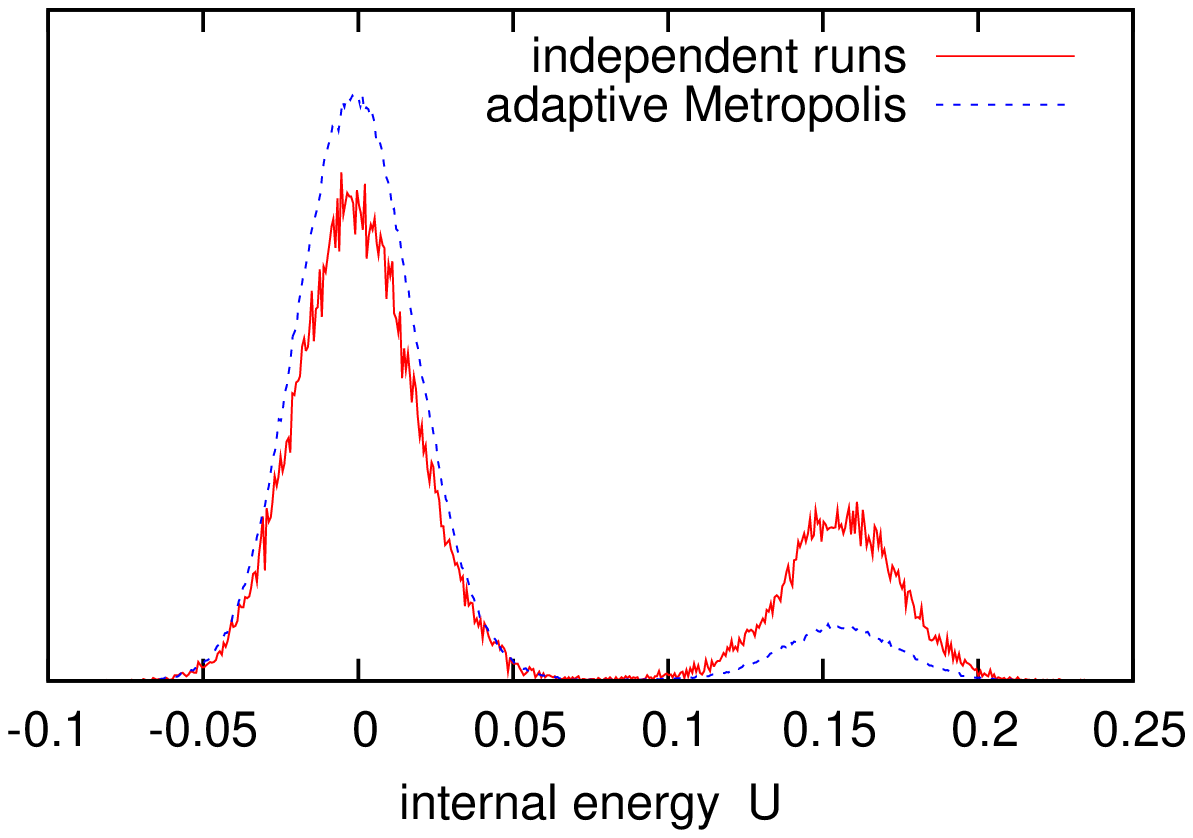}
\hspace*{-2mm}
\includegraphics[angle=270,width=.49\linewidth]{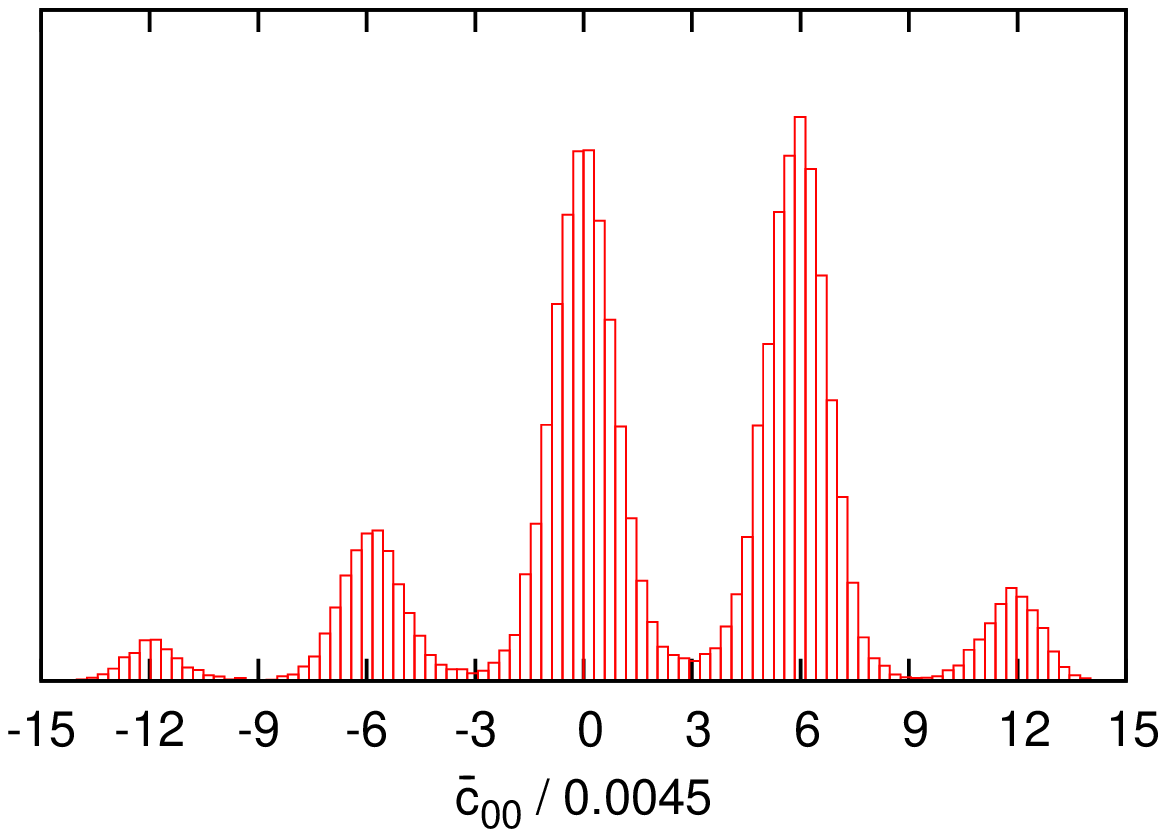}
\vspace*{-2mm} \\
\end{center}
\caption{{\it Histograms for two quantities, which are sensitive to
the variety of action minima.
On the left: the internal energy $U$ measured at $N=12$, $R=4$, 
$\lambda = 1.83$, $m^{2} = -1.7$. 
On the right: $\bar c_{00}$ (see eqs.\ (\ref{clm}),
(\ref{timeav})) measured at $N=12$,
$R=16$, $\lambda = 0.75$, $m^{2} =-0.3$.
$\bar c_{00}$ is shown in units of $2 \sqrt{\pi |m^{2}| / (N^{5/2} \lambda)}
\simeq 0.0045$ so that the peaks are located at integer values,
in agreement with our prediction.}}
\label{histoc00}
\end{figure}

Therefore we extended the algorithm as follows. We stored the
end configuration of each run as $\Phi_{\rm end}$. The subsequent
run takes a new hot start, but after thermalisation the current
configuration may be replaced by $\Phi_{\rm end}$ through a Metropolis
accept/reject step, before the history continues. We denote this
method as {\em adaptive Metropolis algorithm.} In fact it
improves the statistically correct inclusion of the vicinities
around various minima, and it leads to stable and sensible measurements,
as the examples in Figure \ref{metro} illustrate.
The impact of higher local minima tends to be
overestimated by fully independent runs.
The additional Metropolis step helps to overcome this artifact;
in particular the global minimum now receives the suitable weight.
Figure \ref{histoc00} (on the left) shows an example for this 
effect.\footnote{In his study of the 2d model on a fuzzy sphere,
M.\ Panero applied successfully an ``overrelaxation'' technique, 
which is described in his works quoted in Ref.\ \cite{2dfuzzy}.
However, this technique is unlikely to be applicable in our case, 
due to the presence of the temporal kinetic term.} \\

In principle the variety of metastable vacua can be regarded as
a severe thermalisation problem. However, since it is taken care
of by the adaptive Metropolis step, we reduce our notion of
thermalisation to the Monte Carlo time in each run until
the observables stabilise over a long period
at the value that corresponds to the
chosen minimum. In this respect, about 2000 sweeps were
sufficient to thermalise quantities like the action and 
our order parameters (introduced in Section 2).

While this thermalisation is harmless, a technical problem could
occur due to a large number of local minima, in particular
at $\lambda \gg \lambda_{T}$. For a simplified consideration we assume the 
kinetic terms to be negligible (although we saw in Subsection 5.3
that this complete reduction is not really achieved).
Then the matrices $\Phi_{t}$ are independent and --- in a minimum
of the potential --- each one
can be transformed to a diagonal form with elements
$\pm \sqrt{ |m^{2}| / (N \lambda)}$. With all sign combinations
the term $\bar c_{00} = \sqrt{4 \pi} \cdot 
{\rm Tr} (\bar \Phi) / N^{2}$ can take $N^{2} +1$ values,
which is a considerable number for the system sizes that we
studied. However, for independent random signs the values near zero
dominate, whereas the probabilities for minima 
with large $\varphi_{0}= | \bar c_{00}|$ are suppressed. For instance
Figure \ref{histoc00} (on the right) shows a histogram for $\bar c_{00}$
at $N=12\,$; only $5$ peaks (corresponding to the 
$5$ dominant minima) are visible.

The statistical errors were evaluated independently with the binning
and the jackknife method on one hand (we probed various bin sizes),
and with the Madras-Sokal method \cite{MadSok} on the other hand.
The latter amplifies the standard error by a factor which
takes the autocorrelation into account. We generally display
the largest (and therefore safest) error bar obtained by these
methods.

%% file: acknow2.tex
\vspace*{8mm}

\noindent
{\bf Acknowledgements} \ \ We are indebted to Frank Hofheinz for providing
us with a highly optimised parallel code which was applied in this
project. We also thank him, as well as Aiyalam Balachandran,
Brian Dolan, Fernando Garcia Flores, Giorgio Immirzi,
Xavier Martin, Jun Nishimura, Marco Panero, 
Peter Pre\v{s}najder and Jan Volkholz for helpful discussions.
J.M.\ was supported in part by the 
Secretar\'{\i}a de Investigaci\'on y de Posgrado (SIP), and
D.O'C.\ by MTRN-CT-2006-031962.
The simulations were performed on PC clusters at DIAS
in Dublin and at the Humboldt-Universit\"{a}t zu Berlin.